\documentclass[aps,prbbib,floatfix,floats]{revtex4}
\usepackage{graphicx,latexsym}
\usepackage{amsmath}
\usepackage{color}
\usepackage{epsfig}
\begin{document}

\title{Ground state spin and excitation energies in half-filled Lieb lattices}
\author{M. \c Tolea and M. Ni\c t\u a*}

\address{
 National Institute of Materials Physics,
POB MG-7, 77125
 Bucharest-Magurele, Romania.}

\begin{abstract}
We present detailed spectral calculations for small Lieb lattices having up to $N=4$  number of cells, in the regime of
half-filling, an instance of particular relevance for the nano-magnetism
 of discrete systems such as quantum dot arrays,
due to the degenerate levels at mid-spectrum.
While for the Hubbard interaction model -and even number of sites- the ground state spin is given by the Lieb theorem,
the inclusion of long range interaction -or odd number of sites- make the spin state not a priori known, which justifies our approach.
We calculate also the excitation energies, which are of experimental importance,
and find  significant variation induced by the interaction potential.
One obtains insights on the mechanisms involved that impose as ground state
the Lieb state with lower spin rather than the Hund one with maximum spin for the  degenerate levels,
showing this in the first and second order of the interaction potential
for the smaller lattices.
The analytical results concorde with the numerical ones, which are performed by exact diagonalization calculations
or by a combined mean-field and configuration interaction method.
While the Lieb state is always lower in energy than the Hund state,
for strong long-range interaction, when possible, another minimal spin state
is imposed as ground state.
\end{abstract}

\pacs {75.75.-c, 73.22.-f, 71.10.Fd, 71.10.-w}

\maketitle

\section{Introduction}\label{introducere}

The side centered square lattice, i.e. the Lieb lattice, was first proposed in \cite{Lieb} as a rigorous example of itinerant ferromagnetism
in the presence of on-site Hubbard interactions at half-filling.
Recently, the Lieb lattice received renewed attention in the context of
optical and photonic lattices
\cite{Yin, Taie, Apaja, Shen, Mukherjee, Rodrigo},
2D superconductivity \cite{Wang, Iglovikov},
or for its specific topological properties
\cite{Goldman,Weeks,Wei,Palumbo, Jaworowski, Nita}.
In particular, the artificial lattice realization offers the advantage of parameters controlling, leading to various regimes not available in real-atoms lattices, so that one can test a vast spectrum of theoretical prediction.

The Lieb lattice can have non-trivial magnetic properties
\cite{Tamura,Zhao,Chen,Gouveia1,Gouveia2,Xiaodong,Noda1,Noda2}
and its own specificity
originates in the degenerate energy level called also a "flat band" (see, e.g.  \cite{Lieb,Mielke,Tasaki,Derzhko}), which
is located at the middle of the spectrum.
 This flat band is one from the total of {\it three} bands of the Lieb lattice, consistent with the three atoms unit cell. While the electron-hole symmetry imposes that this band is located at precisely {\it zero} energy, its exact degeneracy for a given finite lattice depends also on the border conditions.

 Let us now picture a situation in which such a flat band is half-filled. Then,
a legitimate question would be whether the system obeys the Hund rule with maximum spin
of the electrons on the degenerate levels, say $s=s_{max}$,
or may they have a lower total spin.
The applicability of the Hund rule in various nano-systems have been a subject of considerable interest
(see, e.g. \cite{Steffens,Partoens,Ho,Korkusinski,Karkkainen,Sako,Florens,WSheng,Schroter,Borden}),
both from applicative and fundamental points of view, for understanding the most intimate mechanisms of magnetism.
The results presented in this paper shall add to the existing debate an instance when the Hund rule does not apply.

At this point it is important to mention two well-known theorems that give the ground state spin for some particular lattices with Hubbard interaction. We shall also define some labeling of states used in the paper:
\begin{itemize}
\item {The Lieb theorem} \cite{Lieb} states that, for a half-filled bipartite lattice
(composed of two sub-lattices, say, A and B, and with hopping only between sites from different sub-lattices),
with even total number of sites and with on-site Hubbard repulsion,
the spin of the ground state is
$s_{L}=\frac 12\big| |A|-|B| \big|$
($|A|$, being the number of sites of the sub-lattice A). Needless to say, the Lieb lattice itself is bipartite.
Its ground state spin is thus given by the sites number mismatch between the two sub-lattices,
the theorem stating also that the ground state is not degenerate (excluding the trivial $2s_L+1$ spin degeneracy).
The state lowest in energy from the sub-space with spin $s_L$ shall be referred throughout the paper as the "Lieb" state.
\item {The Mielke theorem} \cite{Mielke} states that a flat band located at the lowest part of the spectrum
will always have the maximum spin ground state $s_{max}$, if filled up to at most half, in the presence of Hubbard interaction.
For exactly half filling, the maximum spin ground state may be degenerate only with the state with a single spin flipped
(excluding again the trivial spin degeneracy).
Throughout this paper, such a state with the maximum spin value of the electrons in a flat band or degenerate level
will be referred as "Hund" state with spin $s_H=s_{max}$, for the correspondence with the atomic physics rule.
\end{itemize}

In this paper we shall address finite Lieb lattices with up to $N=4$ number of cells, or elementary squares, as the one depicted in
Fig.\,1 and we shall impose vanishing boundary conditions. Technically, this means that we can start from the infinite 2D Lieb lattice from which one cuts the smaller lattice of interest
by imposing the wave functions to be {\it zero} on the exterior points  (for the square in Fig.\,1 the wave functions vanish on the sites
$B_{21}$, $B_{22}$, $C_{12}$, etc. -that are not drawn- ), as opposed to periodicity conditions.
The vanishing boundary conditions are particularly relevant for small lattices,
with influence on the physical properties.
In \cite{Tamura} for instance, the authors consider antiperiodic conditions instead, and obtain a different number of levels in the flat band.

For our case, the mid-spectrum level degeneracy is $g=N+1$ \cite{Nita} and
the interesting problem here is that the spin values predicted by the above theorems at half-filling are different.
They are $s_{L}=(N-1)/{2}$ for the Lieb state
and $s_{H}=(N+1)/{2}$ for the Hund one \cite{comment}, being related by the formula $s_L=s_H-1$ and suggesting
a single spin-flip process between them.
However -and as shall be shown- one does not face a contradiction since the Lieb lattices do not have the degenerate flat-band at the bottom of the spectrum, but in its middle,
and we shall show that the interaction with the below electrons proves decisive in imposing the ground state spin.
Nevertheless, it shall be insightful throughout the paper to discuss also the spin properties of
the {\it isolated} degenerate levels for small Lieb lattices, hence the relevance of the Mielke theorem here.

Both the Lieb and Mielke theorems have been rigourously proven only for on-site Hubbard interaction,
so if one includes as well long-range interaction, the results are no longer a priori known, justifying our approach.
Also, for the case of nano-systems, one is typically interested not as much in the ground states configurations,
but especially in the excitation energies   which are the experimentally measurable quantities.
Moreover we shall give insights on the mechanisms that impose the Lieb state as the ground state
using the nonoverlapping property of one of the states in the mid-spectrum \cite{Nita} and the electron-hole symmetry of
the Hamiltonian \cite{Wakabayashi, Nita2, Bogdan}.

In this paper, a particular attention will be paid to the smallest one-cell Lieb lattice depicted in Fig.\,1, which allows (having only eight sites)
an analytical solution at small interactions, as well as numerical exact diagonalization for any value of the interactions.
If only Hubbard interaction is considered  we are in the frame of the Lieb theorem.
However our alternative proof for this specific system allows for an insight on
the role of the states spacial distributions and their symmetry properties.
Similar arguments are presented for two-cell Lieb lattice that, having odd number of sites,
falls outside of the Lieb theorem conditions.
The long-range interaction  included  in our calculations will show that the ground state spin
remains unaffected, even if the electronic configuration itself may change.

For lattices of sizes $N=2\div4$, numerical results will be presented making use of a combined mean-field and configuration-interaction method
(see, e.g.
\cite{Rontani_Chemical,Popsueva,Nielsen,Novak,Schulz,WSheng,Ishizuki,Manolescu,Moldoveanu,Hawrylak,Szafran,Mourad,Odriazola,Rontani_PRL,Ryabinkin,TT,Ferhat}), an approach particularly justified for weak interactions.
The results concur with those obtained for the smallest $N=1$ lattice,
and the Lieb state energy is lower than the Hund state energy both for even and odd number of sites,
even when long range interaction is turned on.
For $N=1$ and $N=2$, the Lieb state \cite{Lieb} corresponds to the minimum spin
[this being $s_L=0(\frac 12)$ for $N=1(2)$], however for  $N=3$ and $N=4$
the "paramagnetic" state -of minimum spin- differs from the Lieb state and emerges as ground state
when the interaction ratio long-range versus Hubbard exceeds a certain value.
This is attributed to the long-range interaction favouring the lowest spin ground state  \cite{LiebM}.

Various shape of nanoscale lattices
can be created as artificial semiconductor quantum dot molecules \cite{Wu}.
{ We briefly mention the experimental realization of
quadruple quantum dots molecules \cite{Romain} and theoretical investigations related
to this subject, including also the interaction effects of half filled systems \cite{teorie1}.
The artificial benzene molecules is theoretically studied \cite{6QD} and is proposed as
an ultracold atom system in \cite{Dirk}.
}
Using GaAs, InAs or Si quantum dots as building blocks
various sizes {of Lieb type systems}
with inter-dot distances $a=5\div 100$\,nm \cite{Tamura} can be tailored.
This opens the posibility to explore the properties of Hubbard like interaction Hamiltonian
in few sites Lieb lattices as studied here.

%
%

The outline of the paper is as follows:
in Section II we give the Hamiltonian and describe the singlet-triplet formulation,
Section III gives both analytical insights and exact diagonalization results for the one cell Lieb lattice,
while Section IV addresses numerically bigger cells with $N=2\div4$.
The Appendixes provide calculations details for the main sections and also analytical insights on the two-cell Lieb lattice.

\section{Interacting Hamiltonian. Singlet and Triplet operators}\label{hint}

Let us consider a 2D lattice with the noninteracting Hamiltonian $\hat H_0$
having the single particle eigenstates { (or noninteracting orbitals in \cite {Benenti})}
$\Phi_\alpha$ and the corresponding energies $\epsilon_\alpha$.
When the electron-electron interaction is considered as well, the total Hamiltonian can be generically written
in the second quantization
\begin{eqnarray}\label{hamiltonian}
\hat H&=&\hat H_{0}+\hat H_{int}=\nonumber \\
&=&\sum_{\alpha,\sigma}\epsilon_{\alpha}c_{\alpha\sigma}^{\dagger}c_{\alpha\sigma}+
\frac12\sum_{\alpha,\beta,\gamma,\delta}\sum_{\sigma\sigma'}V_{\alpha\beta, \gamma\delta}c_{\alpha\sigma}^{\dagger}c_{\beta\sigma'}^{\dagger}
c_{\delta\sigma'}c_{\gamma\sigma},
\end{eqnarray}
where $c^{\dagger}_{\alpha\sigma}$ and its conjugated $c^{}_{\alpha\sigma}$ are the
fermionic creation and annihilation operators of the states $| n_{\alpha\sigma} \rangle$ in the occupation number base,
corresponding to one electron in the state $\Phi_\alpha$ with spin $\sigma=\pm 1/2$.
$V_{\alpha\beta,\gamma\delta}$ are the Coulombian matrix elements  expressed as the scalar products
\begin{eqnarray}\label{potentialabcd1}
V_{\alpha\beta,\gamma\delta}=\langle \Phi_\alpha (1) \Phi_\beta (2) |V(1,2)| \Phi_\gamma(1)\Phi_\delta(2) \rangle,
\end{eqnarray}
with $V(1,2)$ the interaction potential between the particles 1 and 2 { and $\Phi(1)$ or $\Phi(2)$ are
eigenstates of the particle 1 or 2}.
The states and energies of the many-particle Hamiltonian will be noted with $\Psi$ and $E$,
the spin quantum numbers are $s$ for the total spin operator $\hat S$ and $m_s$ for its projection $\hat S_z$.
The energy unit is the hopping integral that is considered $t=1$ and we work with $\hbar=1$.

In the tight-binding model, suitable for lattices such as the Lieb ones we consider here,
the Coulombian matrix elements
are the sum of on-site Hubbard and inter-site long range interaction terms \cite{Hawrylak, Benenti}:
\begin{eqnarray}\label{potentialabcd2}
V_{\alpha\beta,\gamma\delta}=U_H \sum_{i}  \Phi_\alpha(i)^{\ast}\Phi_\beta(i)^{\ast} \Phi_\gamma(i)\Phi_\delta(i)
+V_L\sum_{i\ne j} \frac {\Phi_\alpha(i)^{\ast}\Phi_\beta(j)^{\ast} \Phi_\gamma(i)\Phi_\delta(j)} {|R_i-R_j|},
\end{eqnarray}
where $i,j$ are the discrete lattice sites and $R_i$, $R_j$ their space coordinates that are expressed in terms of the lattice
constant $a$. In Fig.\,\ref{1cell} $a$ is the square length.
$U_H$ and $V_L$ give the Hubbard and long range interaction strengths.
 For a quantum dot array with the confinement potential described in \cite{Tamura}
the Hubbard parameter is $U_H=\frac{\sqrt{2\pi}e^2}{4\pi\epsilon d}$ with $d$ the dot radius depending on the confinement and
$\epsilon$ the dielectric constant.
If we consider the long range interaction parameter $V_L=\frac{e^2}{4\pi \epsilon a}$ we obtain the ratio
$V_L/U_H=\frac{d}{a\sqrt{2\pi}}$.
As example, varying the dot radius $0<d<0.2a$  the ratio $V_L/U_H$ can be modified from 0 to 0.5.
We use these values in the numerical calculations.

As is well known, the Hamiltonian commutes with the spin operators $\hat S$ and $\hat S_z$.
As a consequence, and as will be seen in the following sections,
the eigenfunctions for two electrons will always be singlets ($s=0$, $m_s=0$) or triplets ($s=1$, $m_s=0,\pm1$) states
that are obtained by acting the following singlet and triplet operators on the vacuum:
\begin{eqnarray}
&&\hat S_{\alpha\alpha}=c_{\alpha\uparrow}^{\dagger}c_{\alpha\downarrow}^{\dagger} ,                 \label{singlet2aa} \\
&&\hat S_{\alpha\beta}=\frac{1}{\sqrt 2} \left( c_{\alpha\uparrow}^{\dagger}c_{\beta\downarrow}^{\dagger}
-c_{\alpha\downarrow}^{\dagger}c_{\beta\uparrow}^{\dagger} \right) , ~\text{for}~\alpha \ne \beta,   \label{singlet2ab}\\
&&\hat T_{\alpha\beta}^0=\frac{1}{\sqrt 2} \left( c_{\alpha\uparrow}^{\dagger}c_{\beta\downarrow}^{\dagger}
+c_{\alpha\downarrow}^{\dagger}c_{\beta\uparrow}^{\dagger} \right) ,                                 \label{triplet2ab0}\\
&&\hat T_{\alpha\beta}^{+1}=  c_{\alpha\uparrow}^{\dagger}  c_{\beta\uparrow}^{\dagger} ,            \label{triplet2ab+}\\
&&\hat T_{\alpha\beta}^{-1}=  c_{\alpha\downarrow}^{\dagger}c_{\beta\downarrow}^{\dagger}.           \label{triplet2ab-}
\end{eqnarray}

The singlet and triplet states
are simply a change of basis for the operators pairs that appear in the Hamiltonian, and
we can easily derive the matrix elements of $\hat H_{int}$ in this basis,
relations that will prove useful in the following spectral calculations:
\begin{eqnarray}\label{energiesaasgg}
&&\langle S_{\alpha\alpha}|\hat H_{int}|S_{\gamma\gamma}\rangle= V_{\alpha\alpha,\gamma\gamma} ~,\\
\label{energiesaasgd}
&&\langle S_{\alpha\alpha}|\hat H_{int}|S_{\gamma\delta}\rangle= \sqrt 2 V_{\alpha\alpha,\gamma\delta} ~\text {with}~\gamma\ne \delta, \\
\label{energiesab}
&&\langle S_{\alpha\beta}|\hat H_{int}|S_{\gamma\delta}\rangle= V_{\alpha\beta,\gamma\delta}+V_{\alpha\beta,\delta\gamma} ~
\text {with} ~ \alpha\ne \beta ,~ \gamma\ne \delta ~\text {and} \\
\label{energietab}
&&\langle T_{\alpha\beta}^{m_s}|\hat H_{int}|T_{\gamma\delta}^{m_s}\rangle= V_{\alpha\beta,\gamma\delta}-V_{\alpha\beta,\delta\gamma} ~.
\end{eqnarray}

The full eigenfunctions $\Psi$ for larger number of electrons (for the one-cell lattice for instance we shall need {\it eight} electrons), will be conveniently expressed
by grouping pairs of electrons into singlet and triplet states.

{ We mention that the greek indices $\alpha$, $\beta$ ... are for the single particle eigenstates
(or orbital states) and latin indices $i$, $j$ are for the lattice sites.}

\section{One Cell Lieb Lattice}\label{onecell}

Now we shall specialize the generic Hamiltonian given in the previous Section,
for the particular case of a square with centered sides depicted in Fig.\,\ref{1cell}.
As shall be seen, this smallest realization of a Lieb lattice already has a degenerate level in the middle of the spectrum, raising
non-trivial questions like the ground state spin at half filling or the
values of the first excitation energy.
The two mid-spectrum degenerate levels are spatially disjoint (nonoverlapping),
causing a vanishing exchange interaction between two electrons occupying them and consequently a degeneracy between singlet and triplet states.
We show however that the degeneracy is lift in the favour of the singlet state,
that is remaining the unique ground state when the configurations involving the rest of the spectrum are considered.
We shall present analytical results -considering single electron excitations in the second order of perturbation-
and also exact diagonalization results.

\subsection{Single particle states}\label{sps}

\begin{figure}
\centering
\includegraphics[scale=0.7]{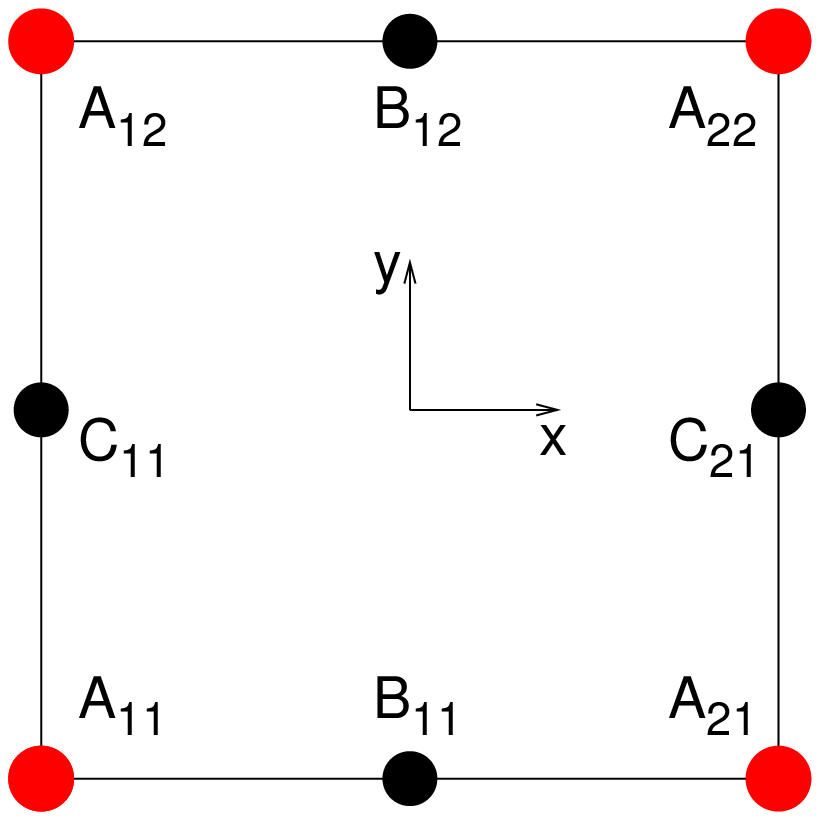}
\includegraphics[scale=0.7]{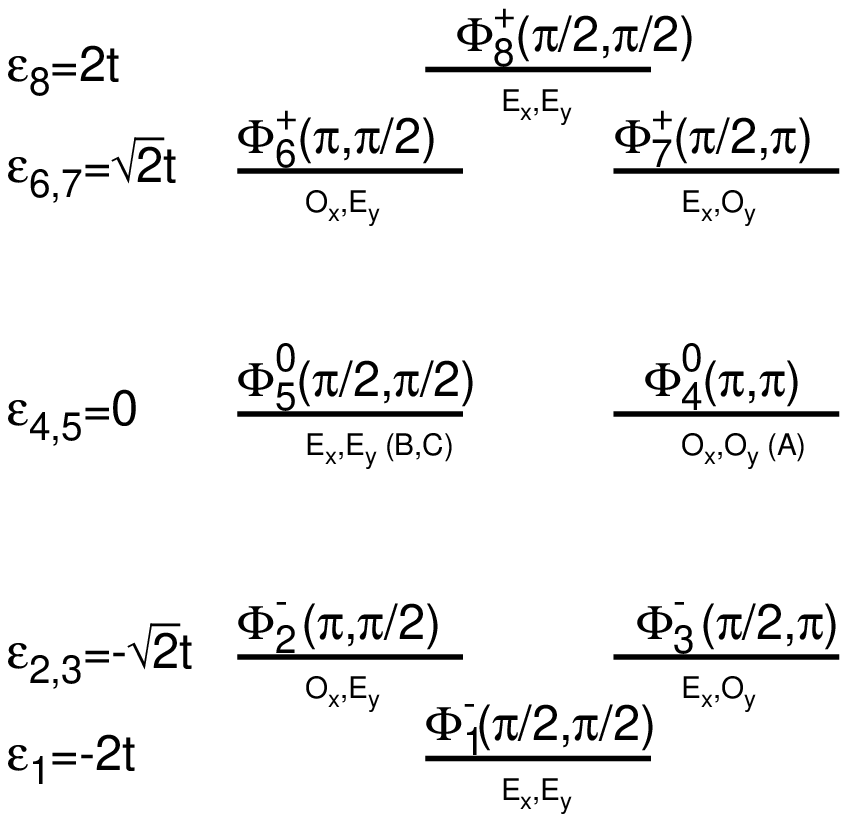}
\caption{(left) The one cell Lieb lattice with $8$ sites.
The indices $n,m$ of the atoms $A$, $B$ and $C$ count the three-sites cells.
(right) The single particle eigenstates $\Phi_\alpha(k_x,k_y)$
and eigenvalues $\epsilon_\alpha$ with $\alpha=1\cdots 8$.
There are three sets of eigenstates: $\Phi^+$ for $\epsilon>0$,
the mid-spectrum degenerate levels $\Phi^0$ for $\epsilon=0$ (called also flat band)
and $\Phi^-$ for $\epsilon<0$. The states marked with $E_x$ and $E_y$ are the states with
even parity on $x$ and $y$ axis respectively and the states marked with $O_x$ and $O_y$ have odd parity.
The two states in the flat band are nonoverlapping being localized on different lattice points, $\Phi_4^0$
on $A$ sites [$\Phi_4^0(A)\ne 0$, $\Phi_4^0(B,C)=0$] and $\Phi_5^0$ on $B, C$ sites [$\Phi_5^0(A)=0$ and $\Phi_5^0(B,C)\ne0$].
}
\label{1cell}
\end{figure}

The one cell Lieb lattice has eight states in the single-particle spectrum as shown in Fig.\,\ref{1cell}.
We shall use the notations from\cite{Nita}, the eight eigenstates being grouped in three branches:
the states $\Phi^{\pm}({\vec k})$
for wave vectors $\vec k=(\pi, \pi/2)$, $(\pi/2, \pi)$ and $(\pi/2, \pi/2)$
with positive (+) and negative (-) energies
$\epsilon^{\pm}({\vec k})=\pm 2t \left({\cos^2 (k_x/2)+\cos^2 (k_y/2)}\right)^{1/2}$
and two states
$\Phi^{0}({\vec k})$ for $\vec k=(\pi, \pi)$ and $(\pi/2, \pi/2)$
with zero energy $\epsilon^0(\vec k)=0$. For simplicity the states are also indexed $\Phi_{1\cdots 8}$
and their energies
are shown in Fig.\,\ref{1cell}.

We give below the expression for the first five quantum states
(vanishing boundary conditions have been implicitly assumed,
meaning that the wave functions are normalized on the eight sites of the system and vanish outside, also no periodicity conditions are imposed):
\begin{eqnarray}
\label{functie1}
&&\Phi^-_1\big(\frac{\pi}{2},\frac{\pi}{2}\big)=\frac{1}{2\sqrt 2}
\big(-|A_{11}\rangle+|B_{11}\rangle -|A_{21}\rangle +|C_{21}\rangle-|A_{22}\rangle
                    +|B_{12}\rangle-|A_{12}\rangle+|C_{11}\rangle   \big), \\
\label{functie2}
&&\Phi^-_2\big(\pi,\frac{\pi}{2}\big)=\frac{1}{2\sqrt 2}
\big(-|A_{11}\rangle +|A_{21}\rangle - \sqrt 2 |C_{21}\rangle + |A_{22}\rangle
                     -|A_{12}\rangle+\sqrt 2 |C_{11}\rangle   \big), \\
\label{functie3}
&&\Phi^-_3 \big( \frac{\pi}{2},\pi \big) =\frac{1}{2\sqrt 2}
\big(-|A_{11}\rangle+\sqrt 2 |B_{11}\rangle -|A_{21}\rangle + |A_{22}\rangle
                    - \sqrt 2 |B_{12}\rangle + |A_{12}\rangle   \big), \\
\label{functie4}
&&\Phi^0_4\big({\pi},{\pi}\big)= \frac{1}{2}
\big(|A_{11}\rangle-|A_{21}\rangle +|A_{22}\rangle -|A_{12}\rangle   \big)~~\text{and} \\
\label{functie5}
&&\Phi^0_5\big(\frac{\pi}{2},\frac{\pi}{2}\big)= \frac{1}{2}
\big(|B_{11}\rangle-|C_{21}\rangle +|B_{12}\rangle -|C_{11}\rangle   \big).
\end{eqnarray}

Some comments are in order:

{\it i.} The states $\Phi^{+}({\vec k})$ are obtained from the states $\Phi^{-}({\vec k})$
changing the sign of A sites localization. This is electron-hole symmetry operation
that change a state with energy $\epsilon$
in the state with energy $-\epsilon$ and change the sign of the wave function projected on one of the sublattices
\cite{Wakabayashi, Nita2, Bogdan}.
In our case $\Phi^{+}(\vec k; A)=-\Phi^{-}(\vec k; A)$ and $\Phi^{+}(\vec k; B,C)=\Phi^{-}(\vec k; B,C)$.

{\it ii.}
In the finite Lieb lattice there is a degenerate level $\epsilon=0$ at mid spectrum
having one of the degenerate states located on A sites
and all of the other states located on B and C sites \cite{Nita}.
For one cell the degeneracy of zero energy level is $g=2$ and,
following the introduced notation, $\Phi_4$ is localized on A lattice sites and
$\Phi_5$ on B, C lattice sites, thus making them spatially disjoint.
This nonoverlapping property of single particle wave functions
one has to keep in mind, as it will play an important role for the many-body spectrum,
leading for instance to the missing of ferromagnetism in a flat band \cite{Patrick}.
The property of certain eigenfunctions being exactly {\it zero} in some lattice sites was proven to be important
also for building generalized eigenfunctions for bigger lattices built by "origami" rules \cite{Dias}.

{\it iii.}
Our Hamiltonian has parity symmetry and
consequently the eigenstates are even or odd in respect to the parity operators $\hat P_x$ and $\hat P_y$
that change $x$ in $-x$ and $y$ in $-y$, respectively.
The states are even at parity operation on $x$ direction when
$\hat P_x \Phi= \Phi$ and we say they have the property $E_x$,
and the states are odd when $\hat P_x \Phi= -\Phi$ and they have the property $O_x$.
For parity on $y$ direction we note with $E_y$ and $O_y$. The parity properties of the eigenstates
are written in Fig.\,\ref{1cell}.
When the electron-electron interaction is considered the parity becomes an important property
because the interaction does not mix  the many-particle states with different parity
due to the selection rules of the Coulombians $V_{\alpha\beta , \gamma\delta}$ defined in Eq.\,\ref{potentialabcd1}.
For instance, an excitation involving an electron transition from the state $\Phi_2^-$ to the state $\Phi_6^+$ is allowed
but to the state $\Phi_7^+$ is forbidden.

\subsection{Two electrons on the degenerate "zero" levels.}\label{2epe2n}

Keeping in mind that the subject of our paper regards the half-filled Lieb lattices
(which for the one cell translates in placing eight electrons on the eight levels),
one can intuitively picture a situation at small interaction strength with the lowest energy states
($\Phi_1$, $\Phi_2$ and $\Phi_3$) double occupied and the remaining two electrons to be placed
on the two degenerate states at mid-spectrum ($\Phi_4$ and $\Phi_5$).
This situation is pictured in Fig.\,\ref{tranzitii}a.

It is instructive to address first the simplified situation in which we neglect the interaction of these two top-most electrons
with the lower fully occupied states,
which is similar to considering an isolated flat band
and places us in the frame of the Mielke theorem \cite{Mielke}.

As the interaction conserves the spin, the Hamiltonian is block diagonal in the total spin subspace,
and the two-particle eigenfunctions are the singlets and the triplets.
The eigenenergies can be straightforwardly derived:
$E(S_{44})=V_{44,44}$, $E(S_{55})=V_{55,55}$, $E(S_{45})=V_{45,45}+V_{45,54}$
and $E(T_{45})=V_{45,45}-V_{45,54}$ with exchange term $V_{45,54}=0$.
Using the single particle functions Eqs.\,\ref{functie4}, \ref{functie5}
one obtains the following energies:
\begin{eqnarray}\label{energys44}
&& E(S_{44})=\frac{4+\sqrt2}{8}{V_L}+\frac{1}{4}U_H \simeq 0.67V_L+0.25U_H,       \\
\label{energys55}
&& E(S_{55})=\frac{1+2\sqrt2}{4} {V_L}+\frac{1}{4}U_H \simeq 0.95V_L+0.25U_H,       \\
\label{energys45}
&& E(S_{45})=\frac{5+\sqrt5}{5} {V_L} \simeq 1.44V_L,       \\
\label{energyt45}
&& E(T_{45}^{m_s})=\frac{5+\sqrt5}{5} {V_L} \simeq 1.44V_L.
\end{eqnarray}

\begin{figure}
\includegraphics[scale=0.6]{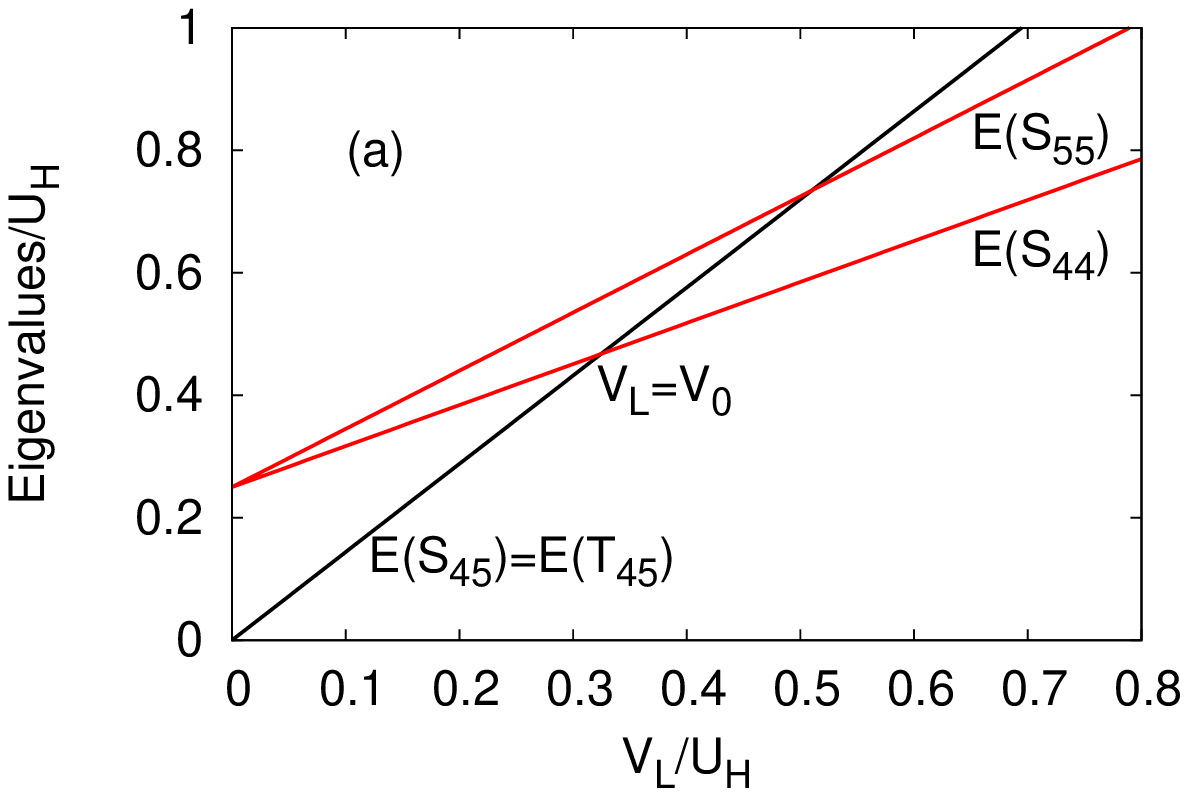}
\includegraphics[scale=0.6]{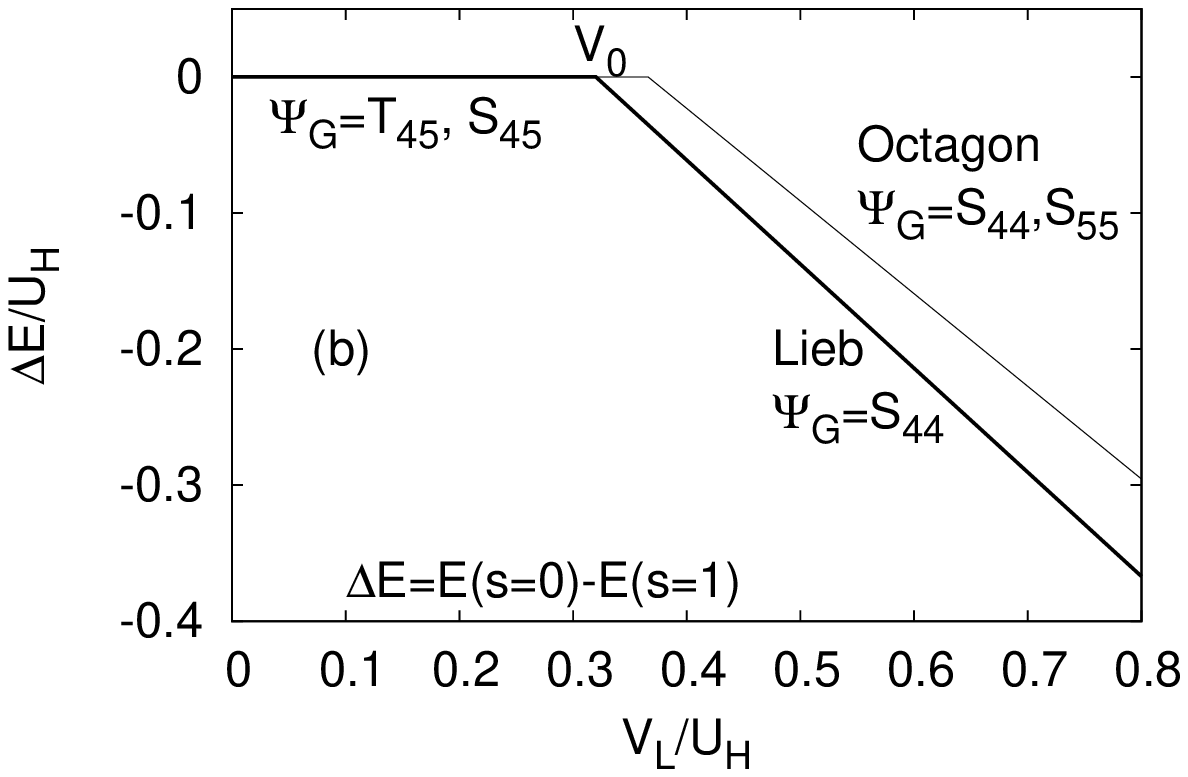}
\caption{(a) Energy evolution of the two-electron states $E/U_H$ from Eqs.\,\ref{energys44}-\ref{energyt45}
versus the ratio $V_L/U_H$ for one-cell Lieb lattice.
For $V_L<V_0$ the degenerate ground state is the singlet state $S_{45}$ and the three triplets $T_{45}^{m_s}$.
For $V_L>V_0$ the role of the long range interaction increases and the nondegenerated ground state is the singlet state $S_{44}$.
(b) The difference between  $s=0$ and $s=1$ ground state energies,
$\Delta E/U_H$, versus the ratio $V_L/U_H$ for one-cell Lieb lattice and for Octagon.
For $V_L<V_0$, $\Delta E=0$ and
for $V_L>V_0$, $\Delta E$ decreases
because a new singlet state become the ground state in $s=0$ spin sector.
The features from (b) exhibit no Hund rule behaviour.
The crossing point in (a) and the sharp decrease of $\Delta E$ in (b) are for the
long range parameter value called $V_0$ that is $V_0\simeq 0.32 U_H$ for Lieb square and $V_0\simeq 0.36 U_H$ for Octagon.
}
\label{evolutievlsiuh}
\end{figure}

Let us briefly discuss the possible ordering on the real axis of the above defined energies.

1. For Hubbard interaction only, $U_H\ne0$ and $V_L=0$, the ground state has the degeneracy $g=4$ and this corresponds to
$E(S_{45})=E(T_{45}^{m_s})\equiv E_G=0$.

2. For long range interaction only, $V_L\ne0$ and $U_H=0$, all energies linearly increase
with $V_L$. The ground state is the singlet state ${S_{44}}$ with $E_G=0.67 V_L$.

3. For both long-range and Hubbard interaction there are two cases that are seen in Fig.\,\ref{evolutievlsiuh}\,(a).
({\it i.})
When $V_L<V_0$ with $V_0 \simeq 0.32U_H$ the ground state has degeneracy g=4,
one is the singlet state $S_{45}$ and three are the triplet states $T_{45}^{m_s}$.
({\it ii.}) When $V_L>V_0$ the ground states is nondegenerated and is the singlet state $S_{44}$.
The term proportional with $V_L$ in formula of $E(S_{44})$ becomes important and
the singlet state $S_{44}$ will have the lowest direct energy of the long range interaction
due to its A site localization (i.e. at the corners of the square).

One can easily calculate the first excitation energy ("demagnetization energy" $\Delta E$) as the difference between the lowest
energy of the nonmagnetic state with total spin quantum number $s=0$ and the lowest energy of the magnetic state with spin $s=1$,
$\Delta E=E(s=0)-E(s=1)$.
We perform this for one cell Lieb lattice and for Octagon with the same electrostatic repulsion between nearby sites
and the results are presented in Fig.\,\ref{evolutievlsiuh}\,(b).
In accordance to the above discussion, one has $\Delta E=0$ for $V_L<V_0$ and a sudden decrease of $\Delta E$ for $V_L>V_0$,
with $V_0\simeq 0.32 U_H$ for one cell Lieb lattice
and $V_0\simeq 0.36 U_H$ for Octagon.
The same qualitative behaviour of sudden decrease for $\Delta E$  at a given ratio $V_L/U_H$ is also seen in
Fig.\,\ref{1cellvu} where the numerical calculations are performed, however at a lower ratio value for Lieb lattice
due to the electrostatic repulsion with the electrons from the other levels
(which supplementary favour the configuration $S_{44}$ with the electrons in the corners being at maximum average distance
from the rest of the charge, as shall be discussed).

For both systems, the one cell Lieb lattice and the Octagon, the calculations in this sub-section show that
the singlet and triplet states are degenerate
for low values of the ratio $V_L/U_H$, while for high values of $V_L/U_H$, the $S_{44}$ singlet becomes the ground state
(degenerated with the $S_{55}$ singlet for the Octagon).
In the next subsection we shall see that the singlet and triplet states degeneracy at low $V_L/U_H$ is lift in the favour
of the singlet state, due to the interaction with the other electrons in the lattice.

{ We mention that the noninteracting eigenspectrum for one-cell Lieb lattice or Octagon
have similar feature with the H\"uckel model \cite{Huckel} for a molecule with eight identical atoms and equivalent bonds.
For instance, our nonoveralapping zero energy eigenstates from Eqs.\,\ref{functie4} and \ref{functie5}
are the well known non-bonding orbitals at the mid-spectrum of planar $D_{8h}$ cyclooctatetraene \cite{COT}.}

\subsection{First and second order approximations.}\label{1cellanalitic}

\begin{figure}
\centering
\includegraphics[scale=0.6]{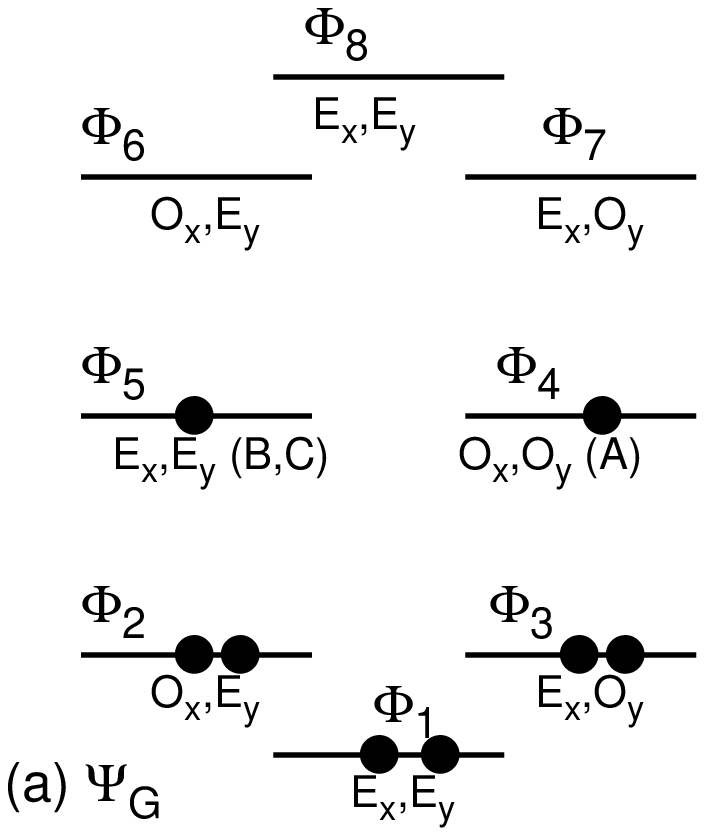}
\includegraphics[scale=0.6]{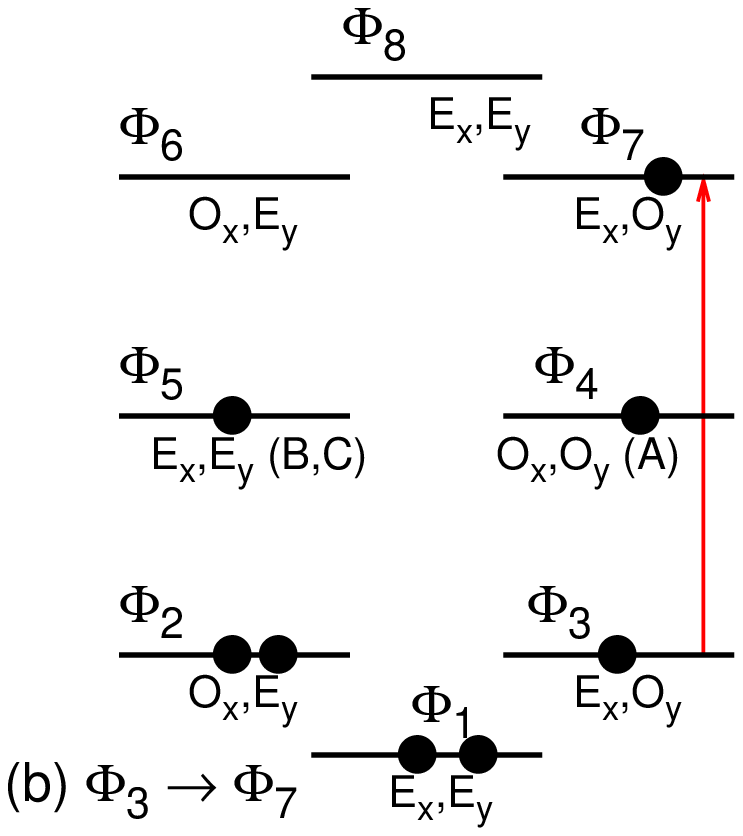}
\includegraphics[scale=0.6]{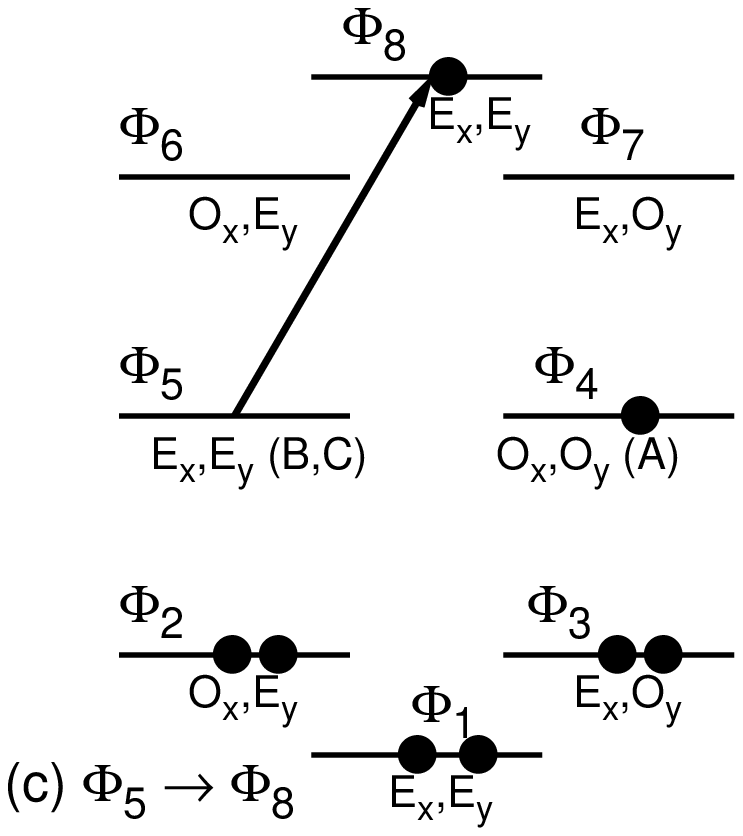}
\caption{The electron configuration of the noninteracting ground state (a) and of the first excited states obtained by single particle excitation
that conserve the parity of the wave functions (b,c)
 for one cell Lieb lattice at half filling.
The configuration in (b) is obtained by a symmetric transition,
$\Phi_3\to \Phi_7$, that change the energy from $-\epsilon_{\vec k}\to \epsilon_{\vec k}$.
The configuration in (c) is obtained by an asymmetric transition $\Phi_5\to \Phi_8$ from a mid-spectrum state to an upper energy state.
The single particle states are explained in Fig.\,\ref{1cell}.}
\label{tranzitii}
\end{figure}

In this sub-section we want to study the lifting of the ground state degeneracy that was seen in the previous subsection
for isolated two degenerate levels if $V_L<V_0$,
and for this purpose one should account for configurations mixing that imply the rest of the spectrum,
by using the full eight electrons wave functions.

The perturbation calculations start with considering the noninteracting ground state, corresponding to
the first six electrons occupying
the lowest single particle states $\Phi_\alpha$ with $\alpha=1,2,3$,
grouped in pairs of singlets $S_{11}$, $S_{22}$ and $S_{33}$.
The last two electrons occupy
the degenerated states $\Phi_4$ and $\Phi_5$ with zero energy, forming pairs of triplet states $T_{45}^{m_s}$
or pairs of singlet states $S_{44}$, $S_{55}$ and $S_{45}$.
In the base of the total spin operators $\hat S^2$ and $\hat S_z$ the six degenerate eigenfunctions and their parity properties are:
\begin{eqnarray}                      \label{states4eu0-1}
&&\Psi_0^{m_s} = S_{11}S_{22}S_{33}T_{45}^{m_s}   ~~~\text{with}~s=1,~m_s=0,\pm 1,~O_x, O_y,    \\
                                      \label{states4eu0-2}
&&\Psi'_0      = S_{11}S_{22}S_{33}S_{45} ~~~\text{with}~s=0,~m_s=0,~O_x, O_y,       \\
                                      \label{states4eu0-3}
&&\Psi''_0     = S_{11}S_{22}S_{33}S_{44}  ~~~\text{with}~s=0,~m_s=0,~E_x, E_y,       \\
                                      \label{states4eu0-4}
&&\Psi'''_0    = S_{11}S_{22}S_{33}S_{55} ~~~\text{with}~s=0,~m_s=0,~E_x, E_y.
\end{eqnarray}
If the interaction is turned on the ground state will be decided between the states $\Psi'_0$ (Lieb) and $\Psi_0^{m_s}$ (Hund),
since the singlets $\Psi''_0$ and $\Psi'''_0$ shall imply highest Coulomb repulsion, as seen in the simplified model
of Fig.\,\ref{evolutievlsiuh} for $V_L<V_0$.

One possible ground state configuration and
two possible configurations obtained by single-particle transitions are shown schematically in Fig.\,\ref{tranzitii}
(spins are not explicitly drawn).

The spin and parity conservation
rules split the total Hilbert space into subspaces,
and only configurations from the same subspace can mix. Relevant subspaces for our discussion  have the spin and parity
properties of the four groups of noninteracting ground states above.

One can define two classes of single-electron transition processes that
conserve both the spin and the parity properties:
({\it i}) First we have the 'symmetric transitions' between states with opposite energies
but the same wave vector $-\epsilon_{\vec k}\to \epsilon_{\vec k}$. In our case they are:
$\Phi_3\to \Phi_7$, $\Phi_2\to \Phi_6$
and $\Phi_1\to \Phi_8$ with general formula of single particle excitation energy $\Delta_{\delta,\gamma}=\epsilon_\delta-\epsilon_\gamma$.
The transition $\Phi_3\to \Phi_7$ is sketched in Fig.\,\ref{tranzitii}b.
As we show below this class of transitions leads to lower energy for the singlet state.
({\it ii}) Second we have 'nonsymmetric transitions' between one state from the flat band to a higher energy state,
$\Phi_5\to \Phi_8$ as sketched in Fig.\,\ref{tranzitii}c or opposite, from a low energy state to the flat band,
$\Phi_1\to \Phi_5$.
However, these nonsymmetric transitions contribute with identical energy shifts for both the singlet and the triplet states
of the pair of flat band electrons (i.e the states $\Psi'_0$ and $\Psi_0^{m_s} $ defined above),
at least up to the second order of our perturbation calculations
(not shown here, as they are technically similar with the ones given below).
As such, they do not contribute to the lifting of the
Lieb-Hund energy degeneracy -our main focus- and can be disregarded at this point.

We present first the calculation for the situation when the possible excited states arise from the electron transition
$\Phi_3\to \Phi_7$. If the other two symmetric transitions are considered the
second energy correction can be shown to be additive.

{\it {\bf 1.} The ground state in the subspace with the total spin $s=1$ and symmetry $O_x$, $O_y$.}
In that case the states have $\hat S_z$ spin degeneracy and we consider the subspace of states
with $m_s=1$.
The situation corresponds to the Hund rule with maximum spin $s_{max}=1$ of the two electrons
on the two degenerate states $\Phi_4$ and $\Phi_5$.
The nonperturbed ground state $\Psi_0$ and three
possible excited states obtained by the single electron transition
$\Phi_3\to \Phi_7$ that conserve the spin and parity properties are:
\begin{eqnarray}
\label{F0}
&&\Psi_0=  S_{11} S_{22} S_{33} T^{+1}_{45}, \\
\label{F1}
&&\Psi_1=  S_{11} S_{22} T^{+1}_{37} S_{45},\\
\label{F2}
&&\Psi_2=  S_{11} S_{22} S_{37} T^{+1}_{45}  ~{\text {and}}\\
\label{F3}
&&\Psi_3= \frac{1}{\sqrt 2} S_{11} S_{22} \left( T^0_{37} T^{+1}_{45}-T^{+1}_{37} T^{0}_{45}   \right),
\end{eqnarray}
with nonperturbed energies $E_0=2\epsilon_1+2\epsilon_2+2\epsilon_3$
and $E_1=E_2=E_3=E_0+\Delta_{7,3}$.
By applying the spin operators $\hat S^2$ and $\hat S_z$
it is readily verified that $\hat S^2 \Psi=2 \Psi $ and $\hat S_z \Psi= \Psi $
meaning $s=1$ and $m_s=1$.

The first order energy correction is
$w_{00}=\langle \Psi_0|\hat H_{int}| \Psi_0 \rangle$ and the second energy correction is given
by the transition amplitudes:
$w_{10}=\langle \Psi_1|\hat H_{int}|\Psi_0\rangle$,
$w_{20}=\langle \Psi_2|\hat H_{int}|\Psi_0\rangle$ and
$w_{30}=\langle \Psi_3|\hat H_{int}|\Psi_0\rangle$.
The first and second order corrections of the energy $E_0$ give the value
\begin{eqnarray}\label{energyEH}
E(s=1)=E_0+w_{00}-\frac{w_{10}^2+w_{20}^2+w_{30}^2}{E_1-E_0},
\end{eqnarray}
with $E_1-E_0=\Delta_{7,3}$.

{\it {\bf 2.} The ground state in the subspace with total spin $s=0$, $m_s=0$ and parity $O_x$, $O_y$.}
This situation corresponds to the ground state spin given by the Lieb theorem
that states the total spin should be $s=0$.
In this case the noninteracting ground state $\Psi'_0$ and two possible
excited states $\Psi'_1$ and $\Psi'_2$
that account for the one electron excitation process
$\Phi_3\to \Phi_7$ are:
\begin{eqnarray}
\label{F0prim}
&&\Psi'_0=S_{11} S_{22} S_{33} S_{45}, \\
\label{F1prim}
&&\Psi'_1=\frac{1}{\sqrt 3}S_{11} S_{22} \left( T^{0}_{37}T^{0}_{45}
          -T^{+1}_{37}T^{-1}_{45} -  T^{-1}_{37}T^{+1}_{45} \right)~{\text {and}}\\
\label{F2prim}
&&\Psi'_2= S_{11} S_{22} S_{37} S_{45},
\end{eqnarray}
with the noninteracting energies $E'_0=E_0$
and $E'_1=E'_2=E_1$ the same as in the subspace with $s=1$.
They are also eigenstates of spin operators with $\hat S^2\Psi=0$ and $\hat S_z\Psi=0$, meaning $s=0$ and $m_s=0$.

The first order energy correction
$w_{00}'=\langle \Psi'_0|\hat H_{int}| \Psi'_0 \rangle$ and the transition amplitudes that give the second energy corrections are
$w_{10}'=\langle \Psi'_1|\hat H_{int}|\Psi'_0\rangle$ and
$w_{20}'=\langle \Psi'_2|\hat H_{int}|\Psi'_0\rangle$.
In the first and second order of the perturbation theory the energy will be
\begin{eqnarray}\label{energyEL}
&&E'(s=0)=E_0+w'_{00}-\frac{{w'}^{2}_{10}+{w'}^2_{20}}{{E'}_1-{E'}_0},
\end{eqnarray}
with ${E'}_1-{E'}_0=\Delta_{7,3}$.

We have derived the following useful relations
for the transition amplitudes:
\begin{eqnarray}
\label{usefullw00}
&&w_{00}^{}=C+ V_{45,45}-V_{45,54},  \\
\label{usefullw00prim}
&&w'_{00}=C+ V_{45,45}+V_{45,54},  \\
\label{usefullw10prim}
&&w'_{10}=\sqrt3 w_{10},         \\
\label{usefullw20prim}
&&w'_{20}=w_{20},
\end{eqnarray}
where the energy term $C$  depends on other interaction processes except for those implying exclusively the
states $\Phi_4$ and $\Phi_5$. The first energy corrections $w_{00}$ and $w'_{00}$
are different by the exchange interaction $V_{45,54}$
as the difference between the simple triplet and singlet states $T_{45}$ and $S_{45}$
that can be seen from Eqs.\,\ref{energiesab} and \ref{energietab}.
The Equations\,\ref{usefullw00},\,\ref{usefullw00prim}, and \ref{usefullw20prim} can be shown by straightforward
calculations and Eq.\,\ref{usefullw10prim} is derived in Appendix \ref{w10siw10p}.

We are interested in the energy difference between the $s=0$ and $s=1$ spin states.
Using  Eqs.\,\ref{energyEH},\,\ref{energyEL}
and relations between the matrix elements, Eqs.\,\ref{usefullw00}...\ref{usefullw20prim},
we calculate that $\Delta E=E'(s=0)-E(s=1)$ depends on $w_{10}$, $w_{30}$, $V_{45,54}$ and does not depend on $w_{20}$.
Consequently we calculate the matrix elements $w_{10}$ and $w_{30}$ obtaining
\begin{eqnarray}                                                \label{calculatw10}
&&w_{10}=\frac{1}{\sqrt 2} \left(V_{74,43}-V_{75,53}\right),\\  \label{calculatw30}
&&w_{30}=     V_{74,43} + V_{75,53},
\end{eqnarray}
and using them
we obtain the following formula for the energy difference $\Delta E$ expressed in the terms on the Coulombian matrix elements and
excitation energy $\Delta_{7,3}=\epsilon_7-\epsilon_3$:
\begin{eqnarray}\label{deltae1}
\Delta E=2V_{45,54}+\frac{4 V_{74,43} V_{75,53}}{\Delta_{7,3}}.
\end{eqnarray}

We remind that $\Delta E$ was obtained considering the single particle excitation $\Phi_3\to \Phi_7$ (Fig.\,\ref{tranzitii}b).
If the others single electron transitions ($\Phi_2\to \Phi_6$ and $\Phi_1\to \Phi_8$) are also considered
the energy difference becomes
\begin{eqnarray}\label{deltae2}
\Delta E=2V_{45,54}+\sum_{(\delta,\gamma)} \frac{4 V_{\delta4,4\gamma} V_{\delta5,5\gamma}}{\Delta_{\delta,\gamma}},
\end{eqnarray}
with the summation over the pairs of the states $(\delta,\gamma)=(6,2)$, $(7,3)$ and $(8,1)$.

{\bf Comment 1.} In the first order of perturbation the singlet and triplet states are still degenerated because
the exchange term $V_{45,54}=0$. It comes from the
nonoverlapping functions $\Phi_4$ and $\Phi_5$ (see Eqs. \ref{functie4} and \ref{functie5}).

{\bf Comment 2.} In the second order of the perturbation and for Hubbard interaction only we have $\Delta E < 0$
meaning the degeneracy is risen and singlet state becomes the ground state. To prove this we
consider the electron hole-symmetry  of the states $(\gamma, \delta)$, meaning that $\Phi_\gamma(B,C)=\Phi_\delta(B,C)$
and $\Phi_\gamma(A)=-\Phi_\delta(A)$, and use the localization properties of states $\Phi_4$ and $\Phi_5$ saying that
$\Phi_4(B, C)=0$ and $\Phi_5(A)=0$.
One obtaines $V_{\delta4,4\gamma}=-V_{\delta5,5\gamma}$ and $\Delta E < 0$.


We show the above result performing the calculation of the interaction matrix elements
in the absence of the long range interaction.
Using the eigenvectors $\Phi^-$ from Eqs.\,\ref{functie1},...,\ref{functie5} and there e-h pairs $\Phi^+$,
from Coulombian matrix Eq.\,\ref{potentialabcd2}
one obtains $V_{\delta 4, 4 \gamma }=-V_{\delta 5,5 \gamma }=-{U_H}/{8}$ for $(\delta,\gamma)=(6,2)$, $(7,3)$ and $(8,1)$.
The single particle excitations energies are $\Delta_{6,2}=\Delta_{7,3}=2\sqrt 2 t$ and $\Delta_{8,1}=4 t$.
The energy difference becomes
\begin{eqnarray}\label{deltae3}
\Delta E=-\frac{U_H^2}{16\sqrt2 t}-\frac{U_H^2}{64 t} \simeq -0.059\cdot \frac {U_H^2}{t}
\end{eqnarray}
that is in good agrement with the parabolic curve obtained in the following numerical calculations of Fig.\,\ref{1cellu}.

This can be regarded as an alternative proof of the Lieb theorem for the particular case of the one-loop Lieb lattice
if only single particle excitation processes are addressed at low interaction.
Supplementary to that, in the low interaction limit, we have proven the missing of linear term and parabolic dependence
of the excitation energy on $U_H$. We notice that one obtains the same conclusions for two cell Lieb lattice
considering the same type of single particle transitions,
and the excitation energy calculation is shortly presented in Appendix \ref{sectiune2celule}.

Numerical calculation using the formula Eq.\,\ref{deltae1} gives negative $\Delta E$ for any ratio $V_L/U_H$.
Finally we remark that
for interaction exceeding a crossing point $V_0$, the formula Eq.\,\ref{deltae1} no longer represents the first excitation energy,
as the ground state in the $s=0$ subspace will be the new singlet $\Psi''_0$ (Eq.\,\ref{states4eu0-3})
this leading to the sharp decrease of $\Delta E$ obtained in the first order of perturbation theory.
This effect is explained in the previous subsection using the two level system.

\subsection{Exact diagonalization}\label{1cellnumeric}

Here we present exact diagonalization results for the half-filled one cell Lieb lattice ($N_e=8$ electrons).
This subsection is complementary to the previous ones, as it does not offer that clear insights on the mechanisms involved,
but on the other hand the results are numerically exact and one is not restricted to small values of the interaction strength ($U_H$ or $V_L$).

\begin{figure}[ht]
\hskip 2cm
\includegraphics[scale=0.44]{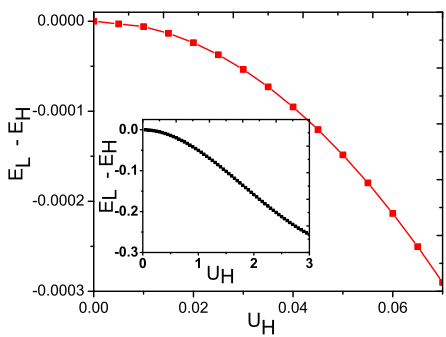}
\vskip -1.5cm
\caption{Exact diagonalization results for the one cell Lieb lattice with Hubbard interaction at half-filling.
The excitation energy between the Lieb ground state (with $s=0$) and the Hund excited state (with $s=1$), $\Delta E=E_L-E_H$,
has a parabolic dependence for small values of $U_H$ with a leading coefficient $\Delta E\simeq -0.059 U_H^2$.
There is a very good concordance with the perturbative analytical results from the previous subsection (see Eq.\ref{deltae3} for low $U_H$ values).
The inset shows that, for larger values of $U_H$ the dependence becomes linear and than sub-linear.}
\label{1cellu}
\end{figure}

In Fig.\,\ref{1cellu} we plot the value of the inverse excitation energy
$\Delta E=E(s=0)-E(s=1)$,
i.e. the  difference between the Lieb ground state energy and the excited Hund state energy, versus $U_H$.
For small values of Hubbard parameter ($U_H<1$) and zero long range interaction, we obtain
the parabolic dependence of $\Delta E$ with the value of the $U_H^2$ leading coefficient $\simeq 0.059$.
This is very close to the analytical one calculated
in the second order of the perturabation theory in Eq.\,\ref{deltae3} for one electron excitation processes.

For larger values of $U_H$, however, the parabolical dependence becomes linear and then sub-linear,
as depicted in the inset of Fig.\,\ref{1cellu}.
For strong interaction energies comparable or exceeding the single-particle level spacing
a very large number of configurations appear in the ground state, including those with two or more electrons on the upper energy levels
(states labeled $\Phi_6$, $\Phi_7$ and $\Phi_8$ in Fig.\,\ref{1cell}).
The analytical insights from Section\,\ref{1cellanalitic}, that are valid for weak interactions, no longer hold.

\begin{figure}[ht]
\includegraphics[scale=0.28]{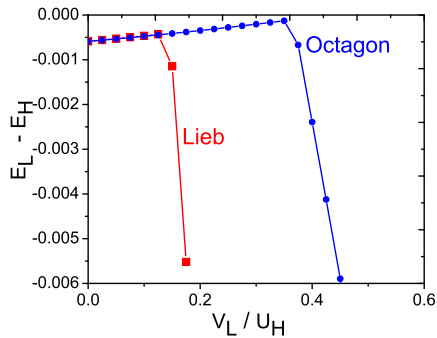}
\vskip -0.5cm
\caption{The energy difference $\Delta E=E_L-E_H$ versus the ratio of the long-range versus Hubbard interactions $V_L/U_H$
for one cell Lieb  lattice and for Octagon at half-filling.
As the ratio $V_L/U_H$ is increased,
one notices first a slight decrease in the excitation energy module
(the difference $\Delta E$ slowly approaches zero),
and then a sharp variation while the ground state changes to a different singlet (see description in text).
The sharp decrease of $\Delta E$ happens at a long range interaction value $V_L=V_0$
that is larger for the Octagon then for the Lieb square.}
\label{1cellvu}
\end{figure}

Next, we see the influence of introducing long-range interaction, an instance for which the distances between points play an important role as well. To illustrate this, we make numerical calculation for two lattice configurations
that  differ by the distances between their points, namely the square shape (Lieb structure) and
Octagon shape (which maximizes the distances between points), the results being shown in Fig.\ref{1cellvu}.

%
%
%
At half-filling both configurations satisfy the Lieb theorem conditions for Hubbard interaction having $s=0$ ground state spin
and we want to see if it changes when the long-range interaction is present.

In Fig.\,\ref{1cellvu} an interesting slight decrease for the excitation energy module, which approaches $zero$, is noticed
as long-range interaction is turned on. This evolution raises the question whether one can induce a ground state spin change
(by the sign change of $\Delta E$), however such an instance was not numerically found neither for Lieb nor for Octagon configurations,
this being in agreement with Eq.\,\ref{deltae2}.

For larger $V_L/U_H$ ratios, one notices a sudden pronounced linear decrease of $\Delta E$
starting from a certain long range parameter value, generically noted with $V_0$,
this being qualitatively similar with the curves obtained for the two level system in Fig.\,\ref{evolutievlsiuh}b.
As described, this is accompanied by the changing of the ground state spatial configuration
and not by the spin change.

The sharp decrease of $\Delta E$ happens at a value of long range parameter that is lower for the Lieb structure
than for the Octagon and below we shall give an insight on why this happens.
For this we have a look at the singlets and triplet energy curves in Fig.\,\ref{evolutievlsiuh}a and
we try to understand how they are changed when the two mid-spectrum electrons start
to interact with the rest of the charge distribution
from the occupied states $\Phi_1$, $\Phi_2$ and $\Phi_3$.

({\it i.}) In the case of a square lattice there is an
energetic advantage for the mid-spectrum electrons to stay in the corners (on state $\Phi_4$)
rather than in the middles of the sides (state $\Phi_5$ is occupied) because this maximizes the distance to the rest of the charge.
The state $S_{44}$ will have a lower increase in energy compared with $S_{45}$ or $T_{45}$
and this can be seen in the difference between diagonal matrix elements of $\hat H_{int}$ for the states $\Psi_0$, $\Psi_0'$ and
$\Psi_0''$ (see Eqs.\,\ref{states4eu0-1}, \ref{states4eu0-2} and \ref{states4eu0-3}).
This will lead to the decrease of the crossing point $V_0$ when the interaction with the rest of the electrons are considered.
(See $V_0\simeq 0.32 U_H$ in Fig.\,\ref{evolutievlsiuh}b and $V_0\simeq 0.14 U_H$ in Fig.\,\ref{1cellvu} for Lieb lattice).
({\it ii.}) This is not the case of an Octagon configuration where the two states
($\Phi_4$ and $\Phi_5$) have equivalent distances to the other
lattice points being only rotated with one site. Consequently the singlet and triplet states energies increases with
equal quantities when the interaction between the two mid-spectrum electrons and the rest of the charge is considered.
It means that the crossing point $V_0$ for the Octagon remain the same at least in the first order of approximation.
(See $V_0\simeq 0.36 U_H$ in Figs.\,\ref{evolutievlsiuh}b and \ref{1cellvu} for Octagon).

\section{Few-cells Lieb lattices. Numerical results}\label{mcellnumeric}

In the previous section we paid a detailed attention to the one cell Lieb lattice,
taking advantage on the fact that the small number of levels allowed both for an analytical insight and for exact diagonalization at half-filling,
however this unfortunately being no longer easy or even possible for bigger lattices.
Here we consider $N=2$, $3$ and $4$ linear cells as the one depicted in Fig.\,\ref{1cell},
meaning lattice dimensions $2\times 1$, $3\times 1$ and $4\times 1$.
The largest  one would require, for instance, placing $23$ electrons on $46$ states for exact diagonalization,
an overwhelmingly demanding computational task (equivalent to diagonalizing a matrix with the size of about $8\cdot 10^{12}$).

We shall address the problem in an approximate manner,
by treating the lowest electrons in a mean-field approximation and the upper ones
(including the ones in the mid-spectrum) by configuration-interaction approach.
{
As in \cite{Rontani_Chemical} the terminology refers to the situation when, even if only a certain number of single-particle levels
 are considered and not all,
(as allowed by the computing power),
importantly, all the Slater determinants for
a given number of electrons
are then built and no truncations are performed in the Fock space.
}
The method works for small interaction strength, when it is justified to treat the lowest occupied states in a mean-field approximation
\cite{Hawrylak}, as the configurations involving high energy single particle excitations have negligible contributions.

The $N$ linear cells Lieb lattice has $5N+3$ single particle states (without spin) and according to
the general counting rule \cite{comment} one have
$N+1$ degenerate states at mid-spectrum
(or flat band) and $2N+1$ states in the upper and lower band respectively.
We shall treat the electrons on the lowest $2N+1$ states in a mean-field approximation, meaning that the modification of the orbitals and of the eigenenergies due to electron-electron interaction is calculated using Hartree-Fock approach and these lowest electrons further influence the remaining higher ones only by the electrostatic potential created. The  configuration-interaction method is then applied for the highest $N+3$ electrons
($N+1$ levels from the flat band plus the first level below and the first one above).

\begin{figure}[h]
\centering
\hskip 2cm
\includegraphics[scale=0.5]{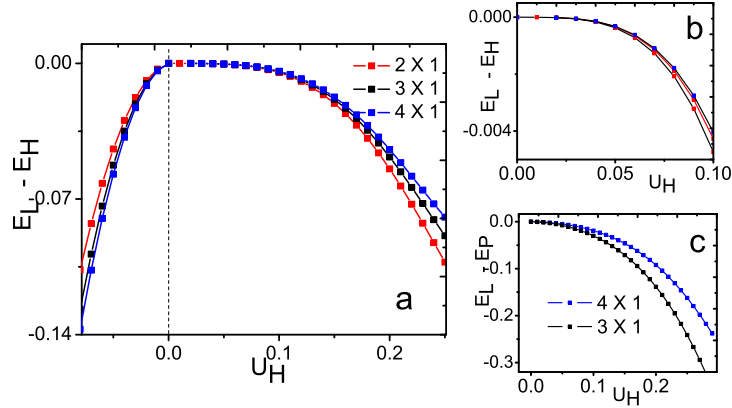}
\vskip -1.5cm
\caption{a) Difference between the Lieb and Hund energies versus $U_H$ for the Lieb lattices of sizes $2\times1$, $3\times 1$
and $4\times1$.
b) A zoom for small $U_H$ to emphasize the parabolic behavior in this range.
c) Difference between the Lieb and Paramagnetic energies, for the $3\times 1$ and $4\times 1$ lattices (when the two energies are distinct). }
\label{mcellu}
\end{figure}

\begin{figure}[h]
\centering
\includegraphics[scale=0.4]{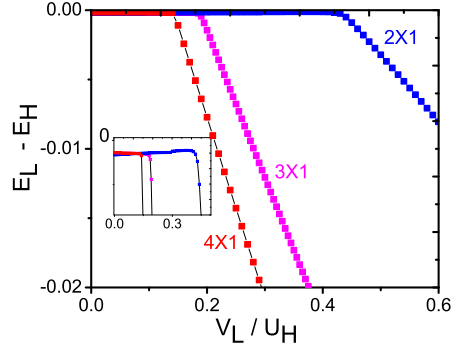}
\vskip -1cm
\caption{Difference between the Lieb and Hund energies as long-range interaction is turned on
(on the $x$ axis one has the ratio between the long range and Hubbard interactions $V_L/U_H$),
for lattices of sizes $2\times 1$, $3\times 1$ and $4\times 1$.
The inset shows a zoom to see that the difference is always negative, the Lieb state being ground state.
The sharp decrease of $\Delta E$ happens at a value of long range called $V_0$ that depends on the lattice size.}
\label{mcellvu}
\end{figure}

\begin{figure}[h]
\centering
\includegraphics[scale=0.30]{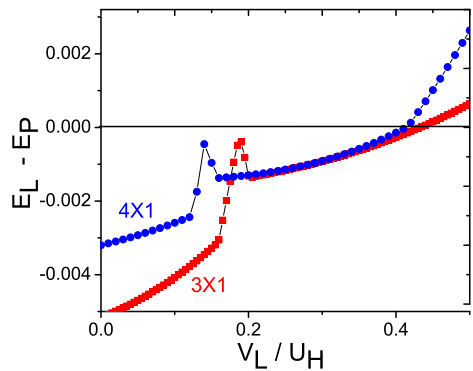}
\caption{Difference between the Lieb ($s_L=s_{max}-1$) and paramagnetic
($s_{min}=s_{max}-2$) energies as function of the ratio $V_L/U_H$
for lattice sizes $3\times 1$ and $4\times 1$.
Notice the non-monotonic behavior (the peak for $V_L\simeq V_0$ with $V_0$ defined in Fig.\ref{mcellvu})
and the sign change (i.e. ground state spin change) for $V_L \simeq 0.4 U_H$.}
\label{mcellepvu}
\end{figure}

Importantly, the lattices of sizes $N=2$ and $N=4$ have odd number of total sites,
which place them outside the strict conditions of the Lieb theorem,
and supplementary motivate our study to determine the ground state spin as well as the excitation energies.

The Hund state have all the electrons
in the flat band with parallel spins,
$s_{H}=s_{max}$ meaning $s_H=(N+1)/2$ \cite{comment}.
We have $s_H=3/2, 2$ and $5/2$ for the cells number $N=2, 3$ and $4$ respectively.
The Lieb state corresponds to the total spin $s_L=\frac 12 (N-1)$
and we have $s_L=1/2, 1$ and $3/2$ for $N=2, 3$ and $4$.
For cells number $N=3$ and $N=4$ three total spin values are possible and
the minimum spin  for them is $s_{min}=0$ and $1/2$ respectively.
They are referred as "paramagnetic" states.

For Hubbard interaction {\it only} we obtain that the Lieb state is the ground state, as for the $N=1$ case discussed in the previous Section,
the difference from the Hund energy being depicted in Fig.\,\ref{mcellu}a.
Fig.\,\ref{mcellu}b is a zoom for small $U_H$, while Fig.\,\ref{mcellu}c shows the energy difference between the Lieb
and the Paramagnetic state.
A parabolic dependence is confirmed for small positive values of $U_H$,
while for negative values of $U_H$ (Fig.\,\ref{mcellu}a) a more abrupt,
almost linear dependence is only noticed. A detailed analysis of the $U_H<0$ regime is not intended here.

Next, we discuss the effect of turning on the long range interaction $V_L$.
We notice the abrupt change of $\Delta E=E_{L}-E_{H}$ for a given value $V_L=V_0$ in Fig.\,\ref{mcellvu}
for similar motifs as discussed for the one-cell case.
We mean that at $V_0$ a new state with a different spatial configuration
becomes the ground state in the spin sector $s=s_L$.
One can argue also that the long-range interaction favours the
lower spin state (Lieb versus Hund).

However, for much higher long range interaction, and if available, the third spin state with even lower spin will have the lowest energy.
This is shown in Fig.\,\ref{mcellepvu}
where the difference between the Lieb state and the Paramagnetic state energies $\Delta E=E_{L}-E_{P}$ is calculated
for the lattices of sizes $3\times 1$ and $4\times 1$.

Before the expected ground state changes into the paramagnetic one,
as the ratio $V_L/U_H$ increases, one notices a non-monotonous dependence around the value $V_L=V_0$
at which the energies in the three different spin sectors $E_H$, $E_L$ and $E_P$ are close together but not equal.

\section{Summary and Conclusions}\label{concluzii}

The Lieb lattices have degenerate energy levels at mid-spectrum, which offer a particular instance to test the alignment of electron
spins in Hund-like situations, and therefore the mechanisms involved in nanomagnetism.
It is shown that the Hund state, with the maximum spin for the electrons on the degenerate levels,
is not the ground state of the half-filled system, the electrons preferring a state with one (or more) spin(s) flipped.

The electron-electron interactions, which are responsible for the non-trivial spin behavior,
have been considered in this work both as on-site Hubbard term ($U_H$)
and as long-range interaction ($V_L$), case which falls outside known theorems such as Lieb \cite{Lieb} or Mielke \cite{Mielke}.
Supplementary, our focus was to calculate excitation energies, which are of experimental relevance.

The special attention is devoted to the smallest lattice shown in Fig.\,\ref{1cell} for two reasons:
({\it i}) it allows both analytical insights in the second order of perturbation theory
and numerical exact diagonalization for the half-filling situation
and
({\it ii}) it already shows the relevant mechanisms we want to discuss.
For the half-filled case, the debate is whether the two top-most electrons on the mid-spectrum degenerate levels
are in the { Hund state} with total spin $s_H=s_{max}$ equal to $1$ in this case (triplet state)
or in the { Lieb state} with $s_L=s_{max}-1$ (singlet state).

We emphasise the wave functions properties
that {\it establish the Lieb state as the ground state}:
({\it a}) one of the states of the mid-spectrum (flat band) have no spatial overlap with the other(s) and
({\it b}) the allowed single particle excitations take place between states related by the electron-hole symmetry
(see for instance the transition in Fig.\,\ref{tranzitii}b).
The property {\it b} systematically leads to negative sign contributions to the level spacing $\Delta E$
between the Lieb and Hund energies, being a second order effect of the interaction potential,
 while the first order correction of $\Delta E$ is zero due to the property {\it a}.
The coefficient of the parabolic dependence of $\Delta E$ on $U_H$ is calculated.

When the long-range interaction is increased and the ratio $V_L/U_H$ exceeds a certain value
one obtains a {\it sharp variation of} $\Delta E$ with no spin change of the ground state.
This is attributed to the crossing between two different singlet states in the many-particle spectrum.
The ground state will change to a different singlet
with two electrons occupying the state located at corners, which minimizes the Coulomb repulsion.
For comparison, it is shown that such a sharp transition occurs for higher values of $ V_L/U_H $
in the case of the Octagon.
The effect is noticed in the first order of perturbation theory and it is confirmed by exact diagonalization calculations,
which allow to address also the case of stronger interactions.

The same two properties {\it a} and {\it b} described above support the Lieb state to have the minimum energy
also for the two cell Lieb lattice with Hubbard interaction, which has odd number of
lattice points placing it outside of the strict Lieb theorem's conditions.

The numerical calculations for $N=2\div4$ size lattices,
using the combined mean-field plus configuration-interaction method,
lead to similar results as the analytical ones, the Lieb state being always lower in energy than the Hund state.
We have again a parabolic dependence of their energy difference  at small $U_H$ (and $V_L=0$)
and we obtain the sharp decrease of the excitation energy while the ratio $V_L/U_H$ exceeds a certain value (for $V_L= V_0$)
that depends on the lattice size.
In the Lieb lattice, there is always one of the degenerate mid-spectrum states
that is nonoverlapping with the others\cite{Nita},
allowing the Lieb state to be degenerate with the Hund one for the {isolated} flat band and to get lower in energy
when configurations involving the rest of the spectrum are taken into account.

For lattices with number of cells $N=3$ and $N=4$
a new state of minimum spin $s_{min}$ has to be considered as possible ground state alongside with the Hund and Lieb states.
Our numerical calculations show that at small long-range interaction the Lieb state remains lower in energy,
while higher values of long-range interaction promotes the $s_{min}$ state as ground state.
In between there is a narrow interval for $V_L$ (around $V_L\simeq V_0$) in which the lowest energies in the three different spin sectors
are very close together (but never quite equal).



The results can be experimental realized in nanoscaled quantum dot devices, artificial molecules or optical lattices
that can be a platform for testing the quantum models as the extended Hubbard Hamiltonian \cite{Tamura, Wu, Walters, Yashwant}.
The interaction parameters $U_H$ and $V_L$ can be tuned for instance by varying the dots sizes, inter-dot distances
or the lattice potential.

One of the main results obtained in the paper is that the few sites Lieb lattice is an interesting example
where the Hund rule of maximum spin does not apply for the mid-spectrum degenerate levels in the presence of Hubbard and long-range interaction as well, giving insightes of the microscopic origin of why this happens.
By varying the interaction strength we obtain two regimes that differ by a strong enhancement of
the Hund state excitation energy when the interacting ratio $V_L/U_H$ exceeds a certain lattice dependent value.

\section{Acknowledgements}
We acknowledge usefull discussions with A. Aldea.
This work is supported by PNII-ID-PCE Research Programme (grant no. 0091/2011) and
the Core Programme.

\appendix

\section{Matrix elements $w_{10}$ and $w^{,}_{10}$}\label{w10siw10p}

In this Appendix we want to prove Equation\,\ref{usefullw10prim}, namely that $w'_{10}=\sqrt 3 w_{10}$.
In the spin subspace with $s=1$, $m_s=1$
we consider the matrix element of $\hat H_{int}$,
\begin{eqnarray}
w_{10}=\langle S_{11} S_{22} T_{37}^{+1} S_{45}  | \hat H_{int} | S_{11} S_{22}  S_{33} T_{45}^{+1}   \rangle.
\end{eqnarray}
In the subspace of $\hat S_z$ spin $m_s=0$ (and not $s=0$) we consider
the matrix elements:
\begin{eqnarray}
&& w_1=\langle S_{11} S_{22} T_{37}^{0} T_{45}^{0}    | \hat  H_{int} | S_{11} S_{22}  S_{33} S_{45}   \rangle, \\
&& w_2=\langle S_{11} S_{22} T_{37}^{+1} T_{45}^{-1}  | \hat  H_{int} | S_{11} S_{22}  S_{33} S_{45}   \rangle, \\
&& w_3=\langle S_{11} S_{22} T_{37}^{-1} T_{45}^{+1}  | \hat  H_{int} | S_{11} S_{22}  S_{33} S_{45}   \rangle.
\end{eqnarray}
We want to show that $w_{10}=w_1=-w_2=-w_3$ that will help us to prove Eq. \ref{usefullw10prim}.

{1.} By applying $\hat S^{-}$ operator on both sides of the scalar product from definition of $w_{10}$
we obtain $w_{10}=\langle S_{11} S_{22} T_{37}^{0} S_{45}  | \hat H_{int} | S_{11} S_{22}  S_{33} T_{45}^{0}   \rangle$.
By writing explicitly $S_{45}$ and $T^0_{45}$ we obtain $w_{10}=w_1$.
We use that any scalar product of $\hat H_{int}$ between vectors that have more than two different occupation numbers
is zero, because $\hat H_{int}$ is biparticle.

{2.} After that we prove that $w_1=-w_2=-w_3$.
For this we start from the expression of $w_1$ and use $T_{37}^{0} T_{45}^{0}=(-1+\frac{1}{2}\hat S^+\hat S^-) T_{37}^{+1} T_{45}^{-1}$.
Considering that the spin operators $\hat S^-$ and $\hat S^+$ commutes with $\hat  H_{int}$
and the action of $\hat S^-$ and $\hat S^+$ on any singlet pair is zero,
we immediately obtain $w_1=-w_2$.

In the same manner, using $T_{37}^{0} T_{45}^{0}=(-1+\frac{1}{2}\hat S^+\hat S^-) T_{37}^{-1} T_{45}^{+1}$ we obtain $w_1=-w_3$.

{3.}
The matrix element $w'_{10}$ is $w'_{10}=\langle \Psi_1'| \hat H_{int}| \Psi_0'\rangle$
with $\Psi_0'$ and $\Psi_1'$ from Eq. \ref{F0prim} and \ref{F1prim}.
We obtain $w'_{10}$ in the terms of $w_1, w_2, w_3$ defined above, $w'_{10}=(w_1-w_2-w_3)/\sqrt 3$.
Using the proved relation $w_{10}=w_1=-w_2=-w_3$  we immediately obtain Equation \ref{usefullw10prim}:  $w'_{10}=\sqrt 3 w_{10}$.

\section{Two cell Lieb lattice. Analytical insights}\label{sectiune2celule}
\begin{figure}
\centering
\includegraphics[scale=0.7]{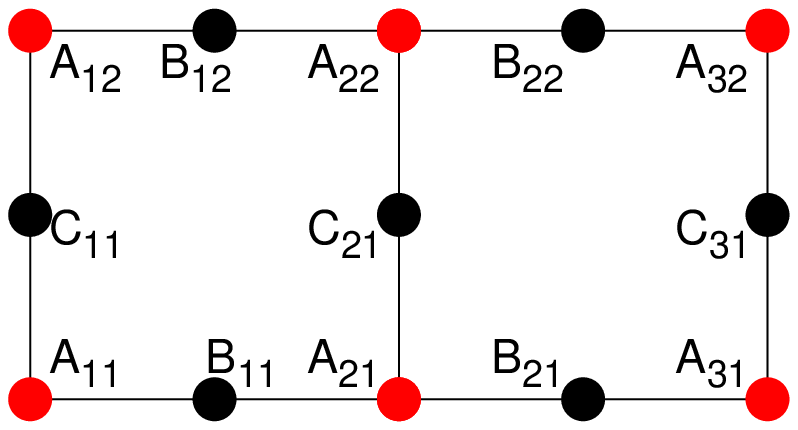}
\includegraphics[scale=0.7]{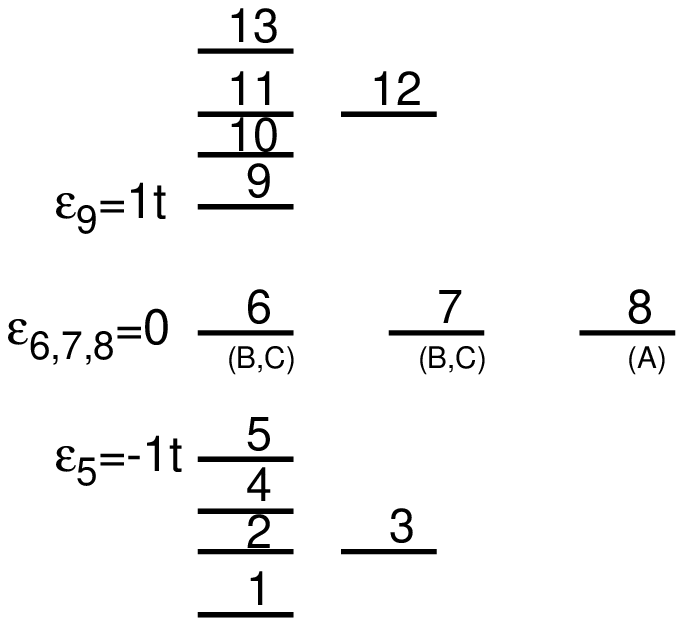}
\caption{(left) The finite two cell Lieb lattice $N=2$ with $13$ sites.
The indices $n,m$ of the atoms $A$, $B$ and $C$ count the unit cells of the lattice.
(right) The single particle eigenstates. Amongst them there are three states with zero energy.
Two of these, $\Phi_6$ and $\Phi_7$, are localized only on $B,C$ sites, and the state $\Phi_8$ has only A sites localization.
}
\label{2cell}
\end{figure}

We consider a two cell finite Lieb lattice ($N=2$) as in Fig.\,\ref{2cell}.
The one electron Hamiltonian has $13$ single particle states, $\Phi_1,\cdots, \Phi_{13}$
with energies schematically depicted in Fig.\,\ref{2cell}\,right. The following remarks can be made:
{\it (i)} The Hamiltonian has the e-h symmetry specific to any bipartite lattice, the eigenvalues being symmetric around $\epsilon=0$.
{\it (ii)} The energy spectrum contains $N+1=3$ {\it zero} energy states with the known localization properties:
two states localized on $B, C$ sites only ($\Phi_6$ and $\Phi_7$)
and one state localized on $A$ sites ($\Phi_8$).
{\it (iii)} Like the one cell finite lattice, the two cell lattice has also parity symmetry,
which can be used as selection rule for the Coulombian matrix elements.

At half filling, for $N_e=13$ electrons, we want to calculate the ground state spin in the second order of approximation
when only one-electron excitation processes are taken into account (symmetric transitions as in Fig.\,\ref{tranzitii}b).
For noninteracting case we consider the many particle ground state
with double occupancy for the lower energy states $\alpha=1,\cdots, 5$
while the zero energy degenerate states $\alpha=6,7,8$ have one electron each of them.
We do not consider the state with double occupancy for $\epsilon=0$ because, when interaction is turned on,
due to Hubbard repulsion, they will have higher energy in the first order approximation already.
We work in the spin subspace with $s=m_s$.
We have three
noninetracting ground states, one with spin $s=3/2$ and two states with $s=1/2$:
\begin{eqnarray}\label{states5eu03pe2}
&&\Psi_{\frac{3}{2}\frac{3}{2}}   =                   S_{11}...S_{55}T_{67}^{+1}n_{8\uparrow}   ,    \\
\label{states5eu01pe2}
&&\Psi_{\frac{1}{2}\frac{1}{2}}   =                   S_{11}...S_{55}S_{67}^{}n_{8\uparrow}   ,    \\
\label{states5eu01pe2prim}
&&{\Psi'}_{\frac{1}{2}\frac{1}{2}}  = \frac{1}{\sqrt 3} S_{11}...S_{55}
     \left(T_{67}^{0}n_{8\uparrow}-\sqrt 2 T_{67}^{+1}n_{8\downarrow}\right)   .    \\
\end{eqnarray}
The symbol $n_{8\sigma}$ means that we have one electron
in the state $\Phi_{8}$ with spin $\sigma$. The spin properties of the states are readily verified by the actions
$\hat S^2 \Psi_{s,m_s}=s(s+1) \Psi_{s,m_s}$ and
$\hat S_z\Psi_{s,m_s}=m_s\Psi_{s,m_s}$.

Following the same steps as for one cell Lieb lattice, we consider the excited states obtained by the one-electron transition
$\Phi_5\to \Phi_9$ that conserve the parity of the system and perform calculation till the second order of approximation.
We give only the principal results.
({\it i}) First we obtain that the interaction potential does not couple between the two $s=1/2$ states, $\Psi_{\frac{1}{2}\frac{1}{2}}$ and
${\Psi'}_{\frac{1}{2}\frac{1}{2}}$,
and the first order lowest energy in their spin sector is for the state ${\Psi'}_{\frac{1}{2}\frac{1}{2}}$.
These are a direct consequence of the different localization of the states $\Phi_8$ and $\Phi_7$ (or $\Phi_6$) and of the fact
that the exchange energy $V_{67,76}$ is a positive quantity, at least for $V_L=0$ (see Eq. \ref{potentialabcd2}),
that makes the triplet state $T_{67}^{m_s}$ to have lower energy than the singlet state $S_{67}$
(see Eqs. \ref{energiesab}, \ref{energietab}).
({\it ii}) Second, the energy difference $\Delta E$ from the ground states with spin $s=1/2$ and with spin $s=3/2$
has the formula
\begin{eqnarray}\label{deltae2c}
\Delta E=\frac{3(V_{56,69} + V_{57,79})V_{58,89}}{\Delta_{9,5}},
\end{eqnarray}
where the first order terms are cancelled out.
The excitation energy is  $\Delta_{9,5}=\epsilon_9-\epsilon_5$. Some brief comments are in order:

{\bf Comment 1.} For Hubbard interaction ($V_L=0$) we have negative $\Delta E$. The states $\Phi_5$ and $\Phi_9$ are related by the e-h symmetry
meaning that $\Phi_5(A)=-\Phi_9(A)$ and $\Phi_5(B,C)=\Phi_9(B,C)$.
For zero energy states we have $\Phi_6(A)=0$, $\Phi_7(A)=0$ and $\Phi_8(B,C)=0$.
From Eq. \ref{deltae2c} and from Coulombian definition Eq. \ref{potentialabcd2}
we obtain that $V_{56,69}$ and $V_{57,79}$ are positive and $V_{58,89}$ is negative.
We immediately have that $\Delta E<0$. It means that the interacting ground state spin is $s=1/2$
this beeing in agreement with the Lieb theorem result
even if it is not strictly applied in our case because the two cell Lieb lattice has an odd number of sites.

{\bf Comment 2.} If we consider other single particle excitation process,  between states
$\Phi_\gamma$ and $\Phi_\delta$ related by e-h symmetry, formula  \ref{deltae2c} is additive.
The cancelation of the first order corrections in the formula of energy difference is preserved and, for Hubbard interaction,
$\Delta E$ remains negative in the second order of perturbation.

\vskip 1cm

* corresponding author e-mail: nitza@infim.ro


\begin{thebibliography}{10}

\bibitem{Lieb} E. H. Lieb, Phys. Rev. Lett. {\bf 62}, 1201 (1989).
\bibitem{Yin} S. Yin, J. E. Baarsma, M.O.J. Heikkinen, J. P. Martikainen, and P. T\"orm\"a, Phys. Rev. A {\bf 92}, 053616 (2015).
\bibitem{Taie} S. Taie, H. Ozawa, T. Ichinose, T. Nishio, S. Nakajima, and Y. Takahashi,
Science Advances 1, 1500854 (2015).
\bibitem{Apaja} V. Apaja, M. Hyrkas, and M. Manninen, Phys. Rev. A {\bf 82}, 041402(R) (2010).
\bibitem{Shen} R. Shen, L. B. Shao, B.  Wang, and D. Y. Xing, Phys. Rev. B {\bf 81}, 041410(R) (2010).
\bibitem{Mukherjee} S. Mukherjee, A. Spracklen, D. Choudhury, N. Goldman, P. \"Ohberg, E. Andersson, and R. R. Thomson,
Phys. Rev. Lett {\bf 114}, 245504 (2015).
\bibitem{Rodrigo}
R. A. Vicencio, C. Cantillano, L. Morales-Inostroza, B. Real,
C. Mej\'ia-Cort\'es, S. Weimann, A. Szameit, and M. I. Molina, Phys. Rev. Lett. {\bf 114}, 245503 (2015).
\bibitem{Wang} H. Wang, S. Yu, and J. X. Li, Phys. Lett. A, 378, 3360 (2014).  
\bibitem{Iglovikov} V. I. Iglovikov, F. Hebert, B. Gremaud, G. G. Batrouni, and R. T. Scalettar,
Phys. Rev. B {\bf 90}, 094506 (2014).                                                  
\bibitem{Goldman} N. Goldman, D. F. Urban, and D. Bercioux, Phys. Rev. A {\bf 83}, 063601 (2011).
\bibitem{Weeks} C. Weeks and M. Franz, Phys. Rev. B, {\bf 82}, 085310 (2010).
\bibitem{Wei} W. F.  Tsai, C. Fang, H. Yao, and J.  Hu, New J. Phys. 17, 055016 (2015).
\bibitem{Palumbo} G. Palumbo and K. Meichanetzidis, Phys. Rev. B {\bf 92}, 235106 (2015).
\bibitem{Jaworowski} B. Jaworowski, A. Manolescu, and P. Potasz, Phys. Rev. B {\bf 92}, 245119 (2015).
\bibitem{Nita}  M. Ni\c t\u a, B. Ostahie, and A. Aldea, Phys. Rev. B {\bf 87}, 125428 (2013).
\bibitem{Tamura}H. Tamura, K. Shiraishi, T. Kimura, and H. Takayanagi, Phys. Rev. B {\bf 65}, 085324 (2002).
\bibitem{Zhao} A. Zhao and S.-Q. Shen, Phys. Rev. B {\bf 85}, 085209 (2012).
\bibitem{Chen} C. Ke-Ji and Z. Wei, Chin. Phys. Lett. 31, 110303 (2014).
\bibitem{Gouveia1} J. D. Gouveia and R. G. Dias, Journal of Magnetism and Magnetic Materials 382, 312 (2015).
\bibitem{Gouveia2} J. D. Gouveia and R. G. Dias, Journal of Magnetism and Magnetic Materials 405, 292 (2016).
\bibitem{Xiaodong} X. Cao, K. Chen, and D. He, J. Phys.: Condens. Matter 27 166003 (2015).
\bibitem{Noda1} K. Noda, A. Koga, N. Kawakami, and T. Pruschke, Phys. Rev. A {\bf 80}, 063622 (2009).
\bibitem{Noda2}
K. Noda, K. Inaba, and M. Yamashita, Phys. Rev. A {\bf 90}, 043624 (2014), Phys. Rev. A {\bf 91}, 063610 (2015).
%
%
%
\bibitem{Mielke} A. Mielke, Phys. Lett. A {\bf 174}, 443 (1993), Phys. Rev. Lett. {\bf 82}, 4312 (1999).
\bibitem{Tasaki} H. Tasaki, Progr. Th. Phys. {\bf 99}, 489 (1998).
\bibitem{Derzhko}O. Derzhko, J. Richter, and M. Maksymenko, Int. J. Mod. Phys. B {\bf 29}, 1530007 (2015).
\bibitem{Steffens} O. Steffens, U. R\"ossler, and M. Suhrke, Europhys. Lett. {\bf 42}, 529 (1998).
\bibitem{Partoens} B. Partoens and F. M. Peeters, Phys. Rev. Lett. {\bf 84}, 4433 (2000).
\bibitem{Ho} A. F. Ho, Phys. Rev. A {\bf 73}, 061601(R) (2006).
\bibitem{Korkusinski} M. Korkusinski, I. P. Gimenez, P. Hawrylak, L. Gaudreau, S. A. Studenikin, and A. S. Sachrajda,
Phys. Rev. B {\bf 75}, 115301 (2007).
\bibitem{Karkkainen} K. K\"arkk\"ainen, M. Borgh, M. Manninen, and S. M. Reimann, New J. Phys. {\bf 9}, 33 (2007).
\bibitem{Sako} T. Sako, J. Paldus, and G. H. F. Diercksen, Phys. Rev. A {\bf 81}, 022501 (2010).
\bibitem{Florens}S. Florens, A. Freyn, N. Roch, W. Wernsdorfer, F. Balestro, P. Roura-Bas, and A. A. Aligia, J. Phys.: Condens. Matter {\bf 23} 243202 (2011).
\bibitem{WSheng}W. Sheng, M. Sun, and A. Zhou, Phys. Rev. B {\bf 88}, 085432 (2013).
\bibitem{Schroter} S. Schr\"oter, H. Friedrich, and J. Madro\~nero, Phys. Rev. A {\bf 87}, 042507 (2013).
\bibitem{Borden} W. T. Borden, H. Iwamura, and J. A. Berson, Acc. Chem. Res. 27, 109-116 (1994).

\bibitem{comment} For a finite Lieb lattice with $N$ cells and vanishing boundary conditions,
the degeneracy of the mid-spectrum level is
$g=N+1$ that is equal to $\big||A|-|B|\big|+2$. The maximum spin at half-filing is $s_H={(N+1)/}{2}$.
The Lieb value corresponds to one spin flipped $s_L= \big||A|-|B|\big|/2= {(N-1)}/{2}$.
In Fig.\,\ref{1cell} the B sub-lattice is labelled both with B and C
to keep the notations in \cite{Nita} for the elementary cell.

\bibitem{Wakabayashi}H. Y. Deng and K. Wakabayashi, Phys. Rev. B {\bf 90}, 115413 (2014).
\bibitem{Nita2} M. Ni\c t\u a, M. \c Tolea, and B. Ostahie, Phys. Status Solidi RRL 08, 790 (2014).
\bibitem{Bogdan} B. Ostahie and A. Aldea, Phys. Rev. B {\bf 93}, 075408 (2016).
\bibitem{Rontani_Chemical}M. Rontani, C. Cavazzoni, D. Bellucci, and G. Goldoni, J. Chem. Phys. {\bf 124}, 124102 (2006).
\bibitem{Popsueva}V. Popsueva, R. Nepstad, T. Birkeland, M. Forre, J. P. Hansen, E. Lindroth, and E. Walterson, Phys. Rev. B {\bf 76}, 035303 (2007).
\bibitem{Nielsen}E. Nielsen, R. P. Muller, and M. S. Carroll, Phys. Rev. B {\bf 85}, 035319 (2012).
\bibitem{Novak}M. P. Nowak, B. Szafran, and F. M. Peeters, J. Phys.: Condens. Matter {\bf 20} 395225 (2008).
\bibitem{Schulz}S. Schulz, S. Schumacher, and G. Czycholl, Phys. Rev. B {\bf 73}, 245327 (2006).
\bibitem{Ishizuki}M. Ishizuki, H. Takemiya, T. Okunishi, K. Takeda, and K. Kusakabe, Phys. Rev. B {\bf 85}, 155316 (2012).
\bibitem{Manolescu}C. Daday, A. Manolescu, D. C. Marinescu, and V. Gudmundsson, Phys. Rev. B {\bf 84}, 115311 (2011).
\bibitem{Moldoveanu}V. Moldoveanu, A. Manolescu, C. S. Tang, and V. Gudmundsson, Phys. Rev. B {\bf 81}, 155442 (2010).
\bibitem{Hawrylak}P. Potasz, A. D. G\"{u}\c{c}l\"{u}, A. Wojs, and P. Hawrylak, Phys. Rev. B {\bf 85}, 075431 (2012).
\bibitem{Szafran} B. Szafran, M. P. Nowak, E. Wach, and D. P. \.{Z}ebrowski, Phys. Lett. A {\bf 378}, 1036 (2014).
\bibitem{Mourad} D. Mourad and G. Czycholl, Eur. Phys. J. B {\bf 78}, 497 (2010).
\bibitem{Odriazola} A. Odriazola, M. M. Ervasti, I. Makkonen, A. Delgado, A. Gonz\'{a}lez, E. R\"{a}s\"{a}nen, and A. Harju, J. Phys: Condens. Matter {\bf 25}, 505504 (2013).
\bibitem{Rontani_PRL} D. Toroz, M. Rontani, and S. Corni, Phys. Rev. Lett. {\bf 110}, 018305 (2013).
\bibitem{Ryabinkin} I. G. Ryabinkin and V. N. Staroverov, Phys. Rev. A {\bf 81}, 032509 (2010).
\bibitem{TT} F. \c Tolea and M. \c Tolea, Physica B {\bf 458}, 85 (2015).
\bibitem{Ferhat} K. Ferhat and A. Ralko, Phys. Rev. B {\bf 89}, 155141 (2014).
\bibitem{LiebM} E. Lieb and D. Mattis, Phys. Rev. {\bf 125} 164 (1962).
%
%
\bibitem{Wu} Jiang Wu, Zhiming M. Wang, Quantum Dot Molecules,
 Lecture Notes in Nanoscale Science and Technology,
 Springer New York, 2013. 
{
\bibitem{Romain} R. Thalineau, S. Hermelin, A. D. Wieck, C. Bauerle,
L. Saminadayar, and T. Meunier, Appl. Phys. Lett. {\bf 101}, 103102 (2012);
M. D. Shulman, O. E. Dial, S. P. Harvey, H. Bluhm, V. Umansky, A. Yacoby,
Science, {\bf 336}, 202 (2012).
\bibitem{teorie1}
%
I. Ozfidan, A. H. Trojnar, M. Korkusinski, P. Hawrylak
Solid State Comm. 172, 15 (2013). 
%
C. Schilling,
Phys. Rev. B {\bf 92}, 155149 (2015).
%
\bibitem{6QD}
G. Begemann, S. Koller, M. Grifoni, and J. Paaske,
Phys. Rev. B {\bf 82}, 045316 (2010);
I. Ozfidan, M. Vladisavljevic,  M. Korkusinski, and P. Hawrylak
Phys. Rev. B {\bf 92}, 245304 (2015).
%
\bibitem{Dirk} D.-S. L\"{u}hmann, C. Weitenberg, and K. Sengstock,
Phys. Rev. X {\bf 5}, 031016 (2015).
}
%
\bibitem{Benenti} G. Benenti, G. Caldara, and D. L. Shepelyansky, Phys. Rev. Lett. {\bf 86}, 5333 (2001).
%
\bibitem{Patrick} P. M\" uller, J. Richter, and O. Derzhko, Phys. Rev. B {\bf 93}, 144418 (2016).
\bibitem{Dias} R. G. Dias and J. D. Gouveia, Scientific Reports {\bf 5}, 16852 (2015).
\bibitem{Huckel} E. H\"uckel, Zeitschrift f\"ur Physik 70, 204 (1931); 72, 310 (1931); 76, 628 (1932); 83, 632 (1933).
\bibitem{COT} C. Trindle and T. Wolfskill, J. Org. Chem. 56, 5426 (1991);
T. Nishinaga, T. Ohmae and M. Iyoda, {\it Symmetry} 2, 76 (2010).
%
%
\bibitem{Walters} R. Walters, G. Cotugno, T. H. Johnson, S. R. Clark, and D. Jaksch, Phys. Rev. A {\bf 87}, 043613 (2013).
\bibitem{Yashwant} Y. Chougale and R. Nath, J. Phys. B: At. Mol. Opt. Phys. {\bf 49}, 144005 (2016). 

\end{thebibliography}
\end{document}